\documentclass[12pt,review,number,sort&compress]{elsart}
\usepackage{times}
\usepackage[dvips]{color}
\usepackage{epsfig}
\usepackage{amsmath}
\usepackage{multirow}
\usepackage{lineno}
 
\begin{document} 

  \linenumbers
\begin{frontmatter}

\title{A Scaler-Based Data Acquisition System for Measuring Parity-Violating Asymmetry in Deep Inelastic Scattering}

\author[0]{R.~Subedi}\footnote{Present address: Richland College, 
Dallas County Community College District, Dallas, Texas 75243, USA},
\author[0]{D.~Wang}, 
\author[2]{K.~Pan},
\author[0]{X.~Deng},
\author[3]{R.~Michaels},
\author[4]{P.~E.~Reimer},
\author[5]{A.~Shahinyan},
\author[3]{B.~Wojtsekhowski},
\author[0]{X.~Zheng\corauthref{cor}}
\address[0]{University of Virginia, Charlottesville, VA 22904, USA}
%\address[1]{George Washington University, 725 21$^{st}$ St, NW, Washington, DC 20052, USA}
\address[2]{Massachusetts Institute of Technology, Cambridge, MA 02139, USA}
\address[3]{Thomas Jefferson National Accelerator Facility, Newport News, VA 23606, USA}
\address[4]{Physics Division, Argonne National Laboratory, Argonne, IL 60439, USA}
\address[5]{Yerevan Physics Institute, Yerevan 0036, Armenia}
\corauth[cor]{Corresponding author.  E-mail: xiaochao@jlab.org; Telephone: 001-434-243-4032; Fax: 001-434-924-4576}

%{\bf{Draft:} \today}
 %\date{\today}

\begin{abstract}

An experiment that measured the parity-violating asymmetries in deep inelastic scattering 
was completed at the Thomas Jefferson National Accelerator Facility in experimental Hall A. 
From these asymmetries, a combination of the quark weak axial charge could be extracted 
with a factor of five improvement in precision over world data. 
To achieve this, asymmetries at the $10^{-4}$ level 
needed to be measured at event rates up to 600~kHz and the high pion background typical to deep inelastic
scattering experiments needed to be rejected efficiently. 
A specialized data acquisition (DAQ) system with intrinsic particle 
identification (PID) was successfully developed and used: The pion contamination in the 
electron samples was controlled at the order of $2\times 10^{-4}$ or below 
with an electron efficiency of higher than 91\% during most of the production period of the experiment,
the systematic uncertainty in the measured asymmetry due to DAQ deadtime was below 0.5\%, 
and the statistical quality of the asymmetry measurement agreed with the Gaussian distribution to 
over five orders of magnitudes. 
The DAQ system is presented here with an emphasis on its design scheme, the achieved PID performance, 
deadtime effect and the capability of measuring small asymmetries. 

\end{abstract}

\begin{keyword}
Jefferson Lab; Hall A; PVDIS; DAQ
\PACS{
%11.    General theory of fields and particles
11.30.Er, %Charge conjugation, parity, time reversal, and other discrete symmetries 
%12.    Specific theories and interaction models; particle systematics
12.15.Mm, %Neutral currents
%12.38.Lg, %Other nonperturbative calculations 
%13.    Specific reactions and phenomenology
13.60.Hb  %Total and inclusive cross sections (including deep-inelastic processes)
%14.    Properties of specific particles
14.60.Cd  %Electrons (including positrons)
14.65.Bt  %Light quarks  
%28.41.Rc Instrumentation -- but this is for Fusion!!!
29.30.Aj %	Charged-particle spectrometers: electric and magnetic 
29.85.Ca %	Data acquisition and sorting 
}% http://www.aip.org/pacs/
\end{keyword}
\end{frontmatter}

%%%%%%%%%%%%%%%%
%\maketitle
%%%%%%%%%%%%%%%%
%\bibliographystyle{apsrev}
%%%%%%%%%%%%%%%%
\section {Introduction}\label{sec:intro}
%%%%%%%%%%%%%%%%

The Parity-Violating Deep Inelastic Scattering (PVDIS) experiment E08-011 was 
completed in December 2009 at the Thomas Jefferson National Accelerator
Facility (JLab). The goal of this 
experiment~\cite{PR08-011,Subedi:spin2008,Paper08-011} was to measure with high precision 
the parity-violating asymmetry in deep inelastic scattering of 
a polarized 6~GeV electron beam on an unpolarized liquid deuterium target. This 
asymmetry is sensitive to the quark weak axial charge $C_{2q}$ which corresponds to
a helicity dependence in the quark coupling with the $Z^0$ boson.

%%%%%%%%%%%%%%%%
For electron inclusive scattering from an unpolarized target, the electromagnetic 
interaction is parity conserving and is insensitive to the spin flip of the incoming 
electron beam. Only the weak interaction violates parity and causes a difference 
between the right- and the left-handed electron scattering cross-sections $\sigma_R$
and $\sigma_L$. The dominant contribution to the parity violation asymmetry, 
$A_{PV}  \equiv ({\sigma_R - \sigma_L})/({\sigma_R + \sigma_L})$, 
arises from the interference between electromagnetic and weak interactions and 
is proportional to the four momentum transfer squared $Q^2$ for $Q^2\ll M_Z^2$. 
The magnitude of the asymmetry is 
on the order of $10^{-4}$ or $10^2$~parts per million (ppm) at $Q^2=1$~(GeV/$c$)$^2$. 

The PVDIS asymmetry from a deuterium target is~\cite{Cahn:1977uu} 
\begin {eqnarray}
A_{PV} =\left(-\frac{G_FQ^2}{4\sqrt{2}\pi \alpha}\right)
  \left(2g_A^e Y_1\frac{F_1^{\gamma Z}(x,Q^2)}{F_1^\gamma(x,Q^2)}+{g_V^e}Y_3\frac{F_3^{\gamma Z}(x,Q^2)}{F_1^\gamma(x,Q^2)}\right)~,
%&&\times\left\{2C_{1u}[1+R_C(x)]-C_{1d}[1+R_S(x)]+\right.\nonumber\\
%&&~~~\left.Y({2C_{2u}-C_{2d}})R_V(x)\right\}~,
\end{eqnarray}
where $Q^2$ is the negative of the four-momentum transfer squared, $G_F$ is the Fermi 
weak coupling constant, $\alpha$ is the fine structure constant, $Y_1$ and $Y_3$ are 
kinematic factors, $x$ is the Bjorken scaling variable, and $F_{1,3}^{\gamma(Z)}(x,Q^2)$ are 
deuteron structure functions that can be evaluated from the parton distribution 
functions and the quark-$Z^0$ vector and axial couplings $g_{V,A}^q$. From this asymmetry 
one can extract the quark weak vector and axial charges $C_{1,2q}$, where the quark weak vector 
charge is defined as $C_{1q}\equiv 2g_A^e g_V^q$ and 
the quark weak axial charge is given by $C_{2q}\equiv 2g_V^e g_A^q$ 
with $q=u,d$ indicating an up or a down quark, 
$g_{A(V)}^e$ is the electron axial (vector) coupling and $g_{V(A)}^q$ is the quark vector (axial) coupling to the $Z^0$ boson. 
In the tree-level Standard Model, the $C_{1,2q}$ are
related to the weak mixing angle $\theta_W$: 
$C_{1u} = -\frac{1}{2} + \frac{4}{3} \sin^2\theta_{W}$, 
$C_{2u} = - \frac{1}{2} + 2 \sin^2\theta_{W}$, 
$C_{1d} = \frac{1}{2} - \frac{2}{3} \sin^2\theta_{W}$, and 
$C_{2d} = \frac{1}{2} - 2 \sin^2\theta_{W}$. 
Although the weak mixing angle and the quark weak vector charge $C_{1q}$ have been measured 
from various processes~\cite{Nakamura:2010zzi}, the current knowledge of the quark weak axial 
charge $C_{2q}$ is poor and their deviations from the Standard Model value would reveal 
possible New Physics in the quark axial couplings that could not be accessed
from other Standard Model parameters.

The goal of JLab E08-011 was to measure the PVDIS asymmetries to statistical precisions
of 3\% and 4\% at $Q^2 = 1.1$ and $1.9$~(GeV/$c$)$^2$, respectively, and under the 
assumption that hadronic physics corrections are small, to extract the quark axial weak 
charge combination $(2C_{2u} - C_{2d})$. In addition, the systematic uncertainty goal was 
less than $3\%$. For this experiment, the expected asymmetries were 91 and 160~ppm 
respectively at the two $Q^2$ values~\cite{PR08-011}.
To achieve the required precision, an event rate capability of up to 600~kHz was needed.

The main challenge of deep inelastic scattering experiments is the separation of
scattered electrons from the pion background in the spectrometer and detector system. 
The neutral pions would decay into $e^+e^-$ pairs from which the electrons produced 
cannot be rejected by detectors. This pair production background was studied by 
reversing the spectrometer magnet settings and measure the $e^+$ yield, and the 
effect on the measured asymmetries was found to be negligible.
Charged pions are produced primarily from nucleon resonance decays and could carry 
a parity violation asymmetry corresponding to the $Q^2$ at which the resonances are produced, 
typically a fraction of the asymmetry of electrons with the same scattered momentum. 
Assuming that a fraction $f_{\pi/e}$ of the detected events are $\pi^-$ and $1-f_{\pi/e}$ are electrons, 
the measured asymmetry is
\begin{eqnarray}
  A_{m} &=& f_{\pi/e} A_\pi + (1-f_{\pi/e}) A_e,
\end{eqnarray}
where $A_e$ is the desired electron scattering asymmetry and $A_\pi$ is the asymmetry
of the pion background. To extract $A_e$ to a high precision, one needs either to minimize
the pion contamination $f_{\pi/e}$ to a negligible level, or to correct the measured asymmetry 
for the asymmetry of pions, which itself needs to be measured precisely. For the PVDIS
experiment, the goal was to control $f_{\pi/e}$ to the $10^{-4}$ level provided that the 
pion asymmetries did not exceed those of electrons.
%Since the expected $\pi$ to electron ratio 
%varies between $(1-10):1$, a $10^4$ pion rejection was needed.

The experiment used a 100~$\mu$A electron beam with a polarization of approximately 
90\% and a 20-cm long liquid deuterium target. The two High Resolution Spectrometers 
(HRS)~\cite{Alcorn:2004sb} were used to detect scattered events. 
While the standard HRS detector package and data acquisition (DAQ) system routinely 
provide a $10^4$ pion rejection with approximately $99\%$ electron efficiency, they 
are based on full recording of the detector signals and are limited to event rates 
up to 4~kHz~\cite{Alcorn:2004sb}.  This is not sufficient for the high rates expected 
for the experiment. (The HRS DAQ will be referred to as ``standard DAQ'' hereafter.)

Recent parity violation electron scattering experiments, %~\cite{Armstrong:2012bi} 
such as %SAMPLE~\cite{Hasty:2001ep} at MIT-Bates, 
HAPPEX~\cite{Aniol:2004hp,Acha:2006my,Aniol:2005zf,Aniol:2005zg,Ahmed:2011vp}, 
and PREX~\cite{Abrahamyan:2012gp} at JLab, focused on elastic scattering from nuclear 
or nucleon targets that are typically not contaminated by inelastic backgrounds. 
Signals from the detectors can be integrated and a helicity dependence in the 
integrated signal can be used to extract the physics asymmetry.  
An integrating DAQ was also used in the preceding PVDIS measurement 
at SLAC~\cite{Prescott:1978tm,Prescott:1979dh} in which approximately 2\% of the 
integrated signal was attributed to pions.  
The SAMPLE experiment~\cite{Hasty:2001ep} at MIT-Bates focused also on elastic scattering but the 
inelastic contamination was more challenging to reject, and an air Cherenkov counter
was used to select only elastic scattering events.
In the Mainz PVA4 experiment~\cite{Maas:2004ta,Maas:2004dh,Baunack:2009gy}, particles were detected 
in a total absorption calorimeter and the integrated energy spectrum was recorded. Charged 
pions and other background were separated from electrons in the offline analysis of the
energy spectrum, and the pion rejection was on the order of 100:1 based on the 
characteristics of the calorimeter. 

High performance particle identification can usually be realized in a counting-based DAQ where
each event is evaluated individually. 
In the G0 experiment~\cite{Beck:1989tg,Armstrong:2005hs,Androic:2009aa,Marchand:2007gc,Androic:2011rha}
at JLab, a superconducting spectrometer with a $2\pi$ 
azimuthal angle coverage was used to detect elastically scattered protons at the forward angle 
and elastic electrons at the backward angle. At the forward angle, protons were identified using 
time-of-flight. At the backward angle, pions were rejected from electrons using an aerogel Cherenkov 
counter, and a pion rejection factor of $125:1$ or better was reported~\cite{Androic:2011rha}. %125:1 is for 400MeV/c pions, the highest momentum range of the backward angle measurement.
The deadtime correction of the counting system was on the order of a few 
percent~\cite{Marchand:2007gc,Androic:2011rha}.
%(G0 instrumentation ref1) and 8-16% for forward angle measurements (G0 instrumentation ref2)

While the PVDIS experiment could fully utilize existing spectrometers and detectors at JLab, 
examination of all existing techniques for PV measurements made it clear that a custom
electronics and DAQ were needed to keep the systematic uncertainties due to data 
collection to below 1\%. 
In this paper we describe a scaler-based, cost effective counting DAQ which limited the pion contamination
of the data sample to a negligible level of $f_{\pi/e}\approx 10^{-4}$. Basic information on the detector
package and the DAQ setup will be presented first and followed by the analysis of electron 
detection efficiency, pion rejection and contamination, corrections due to counting deadtime, 
and the statistical quality of the asymmetry measurement.

%%%%%%%%%%%%%%%%
\section {Detector and DAQ Overview}\label{sec:overview}
%%%%%%%%%%%%%%%%
%
The design goal of the DAQ is to record data up to 600~kHz with hardware-based PID 
and well measured and understood deadtime effects.
The following detectors in the HRS~\cite{Alcorn:2004sb} were used to characterize scattered particles: 
Two scintillator planes provided
the main trigger, while a CO$_2$ gas Cherenkov detector and a double-layer segmented lead-glass 
detector provided particle identification information. The vertical drift chambers 
(as the tracking detector) were used during calibration runs but were turned off
during production data taking because they were not expected to endure
the high event rates. 

For the gas Cherenkov and the lead-glass detector, a full recording 
of their output ADC data was not feasible at the expected high rate. Instead their signals were 
passed through discriminators and logic units to form preliminary electron and pion triggers. 
%Particle identification was fulfilled by the use of discriminators for both the 
%lead-glass and the Cherenkov detectors and proper settings of their thresholds. 
These preliminary triggers were then combined with the scintillator triggers  
to form the final electron and pion triggers, which were
sent to scalers to record the event counts and used offline to form 
asymmetries $A = (n_R - n_L)/(n_R + n_L)$, where $n_{R(L)}$ is
the integrated rate of the triggers normalized to the integrated beam charge 
for the right$(R)$ and left$(L)$ handed spin (helicity) states
of the incident electron beam.  
The scalers that counted triggers and the beam charge were integrated over the helicity 
period, which was flipped pseudo-randomly at 30 Hz per the 
experimental technique used by the HAPPEX experiments~\cite{Ahmed:2011vp}.

For the HRS the two layers of the lead-glass detector are called ``preshower'' and ``shower'' detectors, 
respectively. The preshower in the Right HRS (the spectrometer located to the right 
side of the beamline when viewed along the beam direction) has $48$ blocks arranged in a $2\times 24$ 
array, with the longest dimension of the blocks aligned perpendicular to the particle trajectory. For the two blocks
in each row, only the ends facing outward are read out by photo-multiplier tubes (PMTs), while 
the other ends of the two blocks are facing each other and not read out. Therefore, the 
preshower detector has $48$ output channels. All preshower blocks were individually wrapped 
to prevent light leak. The shower detector in the Right HRS had $75$ blocks 
arranged in a $5\times 15$ array with the longest dimension of the blocks aligned 
along the trajectory of 
scattered particles. PMTs were attached to each block of the Right shower detector on one end only, 
giving normally $75$ output channels. However to minimize the electronics needed for this experiment 
(see next paragraph), 
only 60 of the 75 shower blocks were used
while signals from the 15 blocks on the edge were not utilized by the DAQ. The reduction
of the HRS acceptance due to not using these side blocks was negligible.
The preshower and the shower detectors in the Left HRS are similar to 
the preshower detector on the Right HRS except that for each detector there are $34$ blocks 
arranged in a $2\times 17$ array.

\begin{figure}[!ht]
\hspace*{0.6cm}
\includegraphics[width=0.36\textwidth]{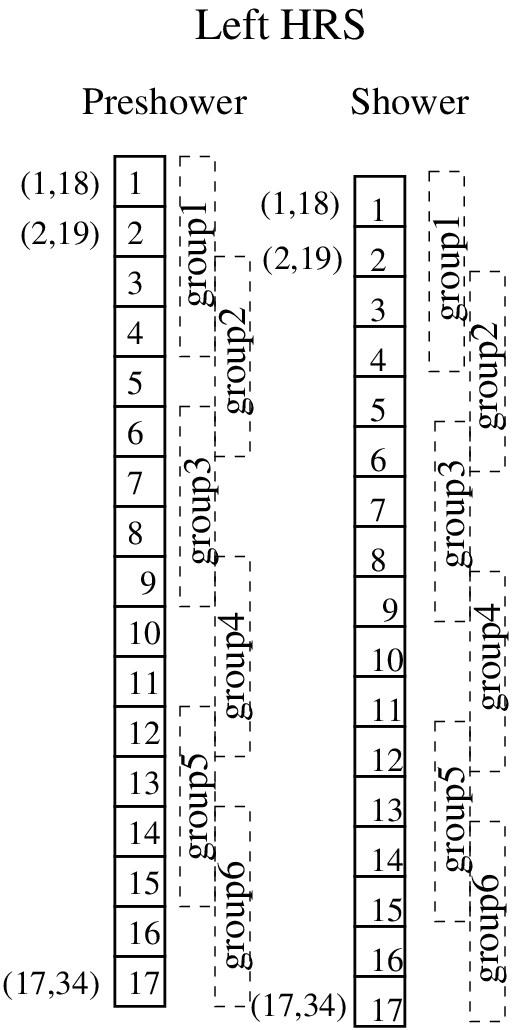}
\hspace*{0.6cm}
\includegraphics[width=0.54\textwidth]{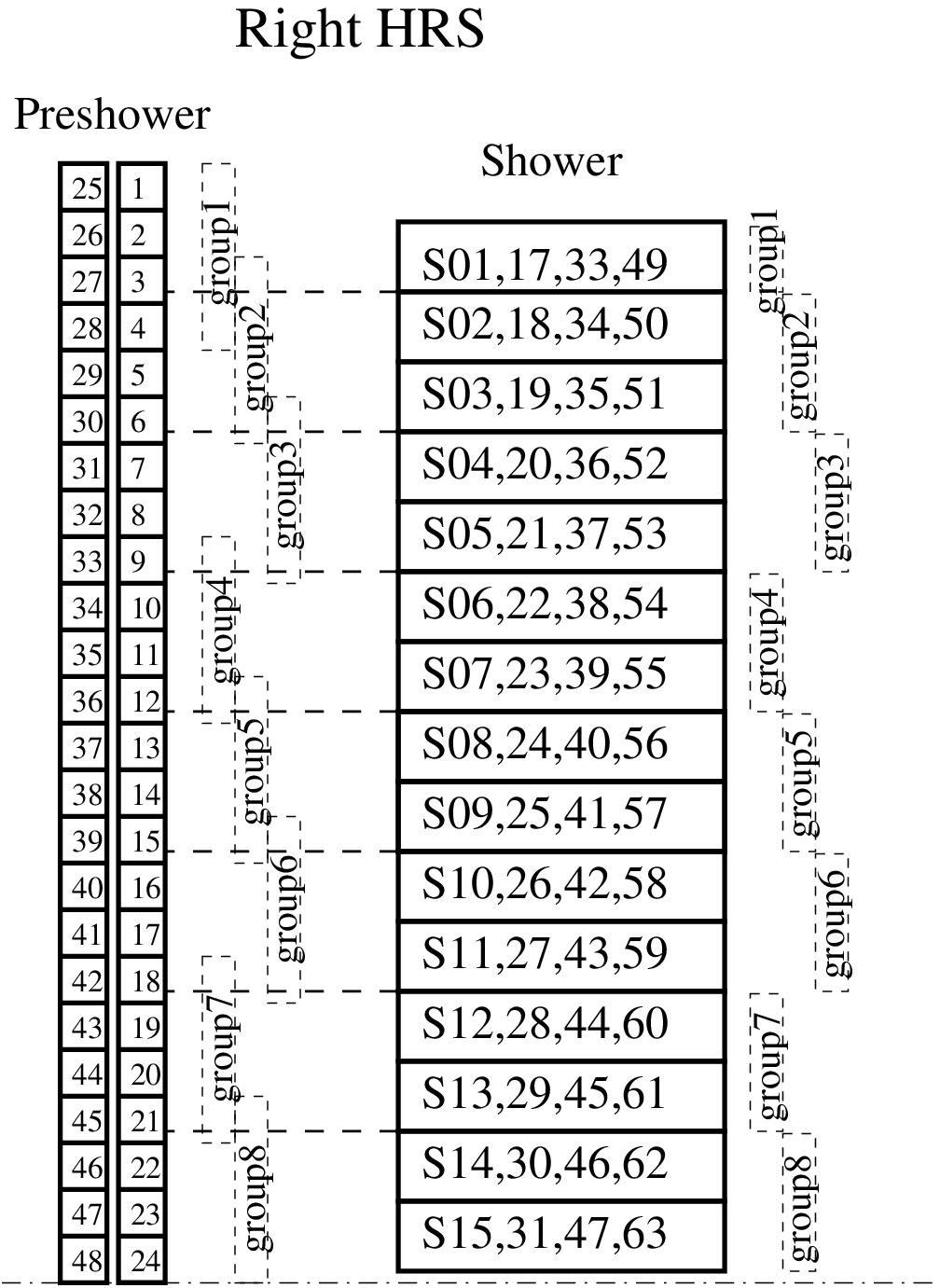}
\caption{Grouping scheme (side-view) for the double-layer lead-glass detectors 
for the Left and the Right HRS. Scattered particles enter the detector
from the left. The dashed vertical bars represent the range of each group.
The Right HRS Shower blocks are labeled as 1 through 64 for historical reasons, 
but row 16 (blocks 16, 32, 48 and 64) was not present during this experiment.
}\label{fig:grouping}
\end{figure}

Because the lead-glass detectors in the Left and Right HRS are different, design of 
the lead-glass-based triggers of the DAQ is also different, as shown in Fig.~\ref{fig:grouping}.
As a compromise between the amount of electronics needed and the rate in the 
front end logic modules, 
the lead-glass blocks in both the preshower and the shower detectors were divided into 6 (8) groups 
for the Left (Right) HRS, with each group consisting typically 8 blocks. 
Signals from the 8 blocks in each group were added using a custom-made 
analog summing unit called the ``SUM8 module'', then passed to discriminators. The geometry and the position of 
each preshower group were carefully chosen to match those of the corresponding shower group 
to maximize electron detection efficiency.  On the Left HRS, adjacent groups in both preshower 
and shower had overlapping blocks, while for the Right HRS only preshower groups were 
overlapping. 
To allow overlap between adjacent groups, signals from preshower blocks on the Right HRS
and from both preshower and shower blocks on the Left HRS were split into two identical copies
using passive splitters. 

\begin{figure*}%% \begin{figure*} and \end{figure*} allows to fit the figure in both columns in a page with 2-columns.
%color version
\includegraphics[width=1.0\textwidth,angle=0]{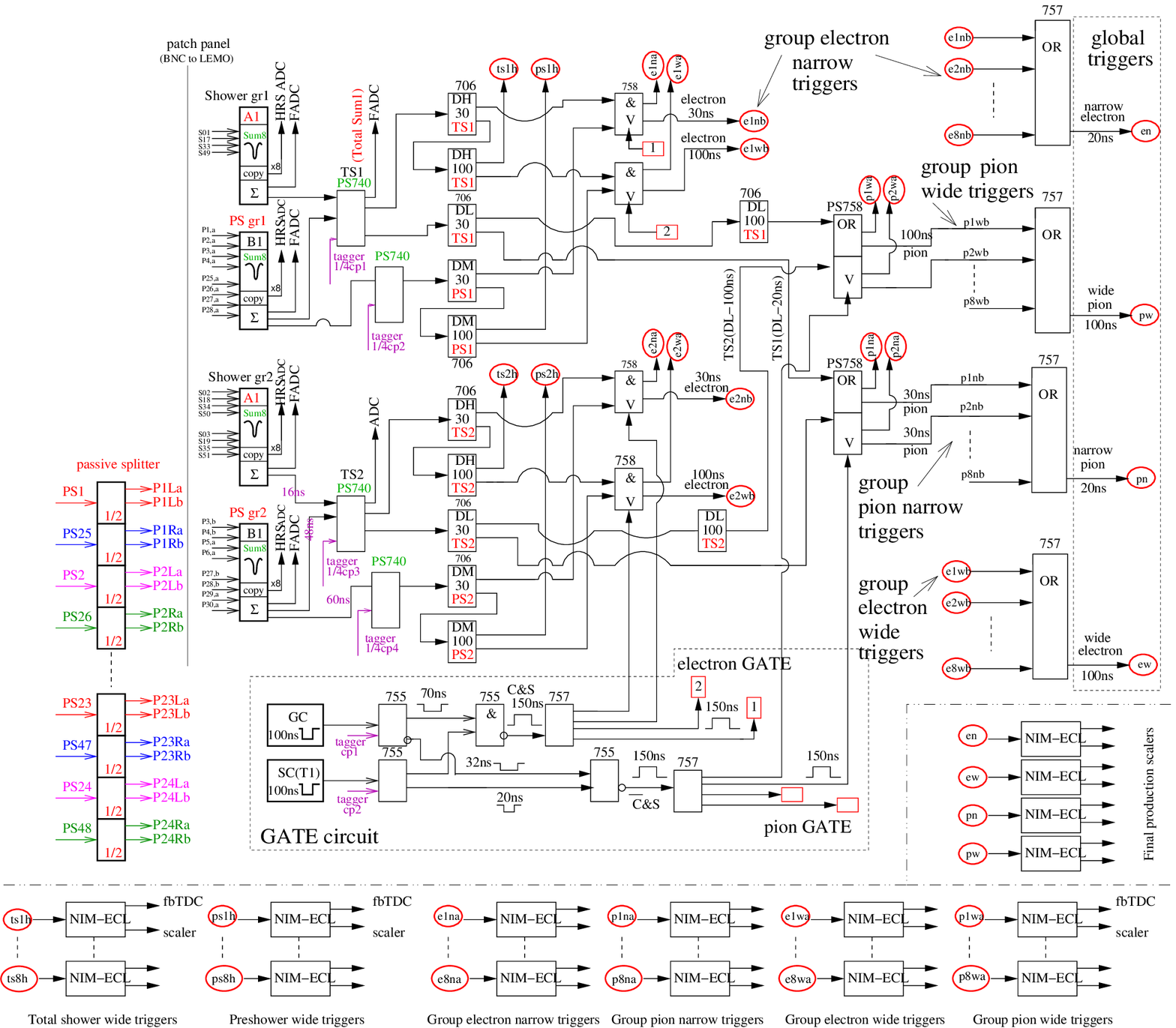}
%black&white version
%\includegraphics[width=1.0\textwidth,angle=0]{Parity_trig3R_March2013_forNIM_bw.eps}
\caption{[{\it Color online}] Electronics diagram for the Right HRS DAQ used by the PVDIS experiment.
The Sum8's, discriminators and logic modules for two groups are shown, as well
as the location of tagger signal inputs, setup of the GATE circuit using 
scintillator (SC) and gas Cherenkov (GC) signals, the logic units for combining triggers 
from all eight groups into final triggers, the counting scalers, and the monitoring fastbus
TDCs. The discriminators had three different levels of threshold settings: low threshold 
(DL) was used on the total shower (TS) signals to form pion triggers; medium (DM) and high 
thresholds (DH) were used on preshower (PS) and TS signals respectively to form electron 
triggers. During the experiment the thresholds were approximately -20~mV for DL and DM, 
and in the range $(-50,-70)$~mV for DH depending on the momentum setting of the spectrometer. 
Electronics for the Left HRS are similar except for the grouping scheme.
}
\label{fig:daqflowchart}

\end{figure*}

A schematic diagram of the DAQ electronics for the Right HRS is shown in 
Fig.~\ref{fig:daqflowchart}. Preliminary electron and pion triggers were formed by passing 
shower (SS) and preshower (PS) signals and their sums, called total shower (TS)
signals, through discriminators with different thresholds. 
For electron triggers, logical ANDs of the PS discriminator and the TS
discriminator outputs were used. For pions, low threshold discriminators on
the TS signal alone were sent to logical OR modules to produce preliminary triggers. 
Additional background rejection was provided by the ``GATE'' circuit, which combined
signals from the gas Cherenkov (GC) and the ``T1'' signal~\cite{Alcorn:2004sb} from the scintillators (SC). 
Each valid coincidence between GC and T1 would produce a 150-ns wide electron GATE
signal that allowed an output to be formed by the logical AND modules from the preliminary 
electron triggers. Each valid T1 signal without the GC signal would produce a 150-ns
wide pion GATE signal that allowed an output to be formed by the logical OR modules
from the preliminary pion triggers. The outputs of the logical AND and OR modules are
called group electron and pion triggers, respectively. 
All six (eight) group electron or pion triggers were then ORed together 
to form the global electron or pion trigger for the Left (Right) HRS. 
All group and the final electron and pion triggers were counted using scalers. 
Because pions do not produce large enough lead-glass signals to trigger the
high threshold TS discriminators for the electron triggers, pions do not introduce extra
counting deadtime for the electron triggers. However, the 150-ns width of the electron 
GATE signal would cause pion contamination in the electron trigger. This effect will be
presented in Section~\ref{sec:pid}.

In order to monitor the counting deadtime of the DAQ, two identical paths of
electronics were constructed. The only difference between the two paths is in the PS and the TS
discriminator output widths, set at 30~ns and 100~ns for the ``narrow''
and the ``wide'' paths, respectively. 
The scalers are rated for 250 MHz (4~ns deadtime) and therefore do not add 
to the deadtime.  In addition, the output width of all logic modules was set to 15~ns, 
so the deadtime of the DAQ for each group is dominated by the 
deadtime of the discriminators. Detailed analysis of the DAQ deadtime will be presented in 
Section~\ref{sec:deadtime}.

The SUM8 modules used for summing all lead-glass signals also served as fan-out 
modules, providing exact copies of the input PMT signals. These copies were sent 
to the standard HRS DAQ for calibration. 
%, hence the standard DAQ remained fully functional. 
During the experiment, data were collected at low rates using reduced beam currents 
with both DAQs functioning, such that a direct comparison of the two DAQs could be made. 
Vertical drift chambers were used during these low rate DAQ studies. 
Outputs from all discriminators, signals from the scintillator and the gas Cherenkov,
and all electron and pion group and global triggers were sent to Fastbus TDCs (fbTDC) and were recorded 
in the standard DAQ. Data from these fbTDCs were used to align the amplitude spectrum
and timing of all signals. They also allowed the study of the Cherenkov and the 
lead-glass detector performance for the new DAQ.

Full sampling of partial analog signals was done using Flash-ADCs (FADCs) 
at low rates intermittently during the experiment. For one group on the Left and one 
group on the Right HRS, the preshower and the shower SUM8 outputs, the intermediate logical
signals of the DAQ, and the output electron and pion triggers were recorded. 
These FADC data provided a study of pileup effects to confirm the deadtime simulation and to
provide the input parameters for the simulation, specifically the rise and fall times
of the signals and their widths.

\section{Overview of Kinematics}\label{sec:exp_overview}
During the experiment data were taken 
at two deep inelastic scattering (DIS) kinematics at $Q^2=1.1$ and $1.9$~(GeV/$c$)$^2$. 
These were the main production kinematics and will be referred to as 
DIS\#1 and DIS\#2, respectively. Due to limitation of the spectrometer magnets, DIS\#1 was
taken only on the Left HRS, while DIS\#2 was taken on both Left and Right
HRSs.
In addition, data were taken at five kinematics within or near the nucleon resonance 
region with their invariant mass $W$ between the $\Delta(1232)$ resonance and just above $W=2$~GeV. 
These data were used for the purpose of radiative corrections and will be referred to as
RES I through V (although kinematics V was located slightly above $W=2$~GeV). Data for each of 
the resonance settings were taken only with one HRS because of the spectrometer magnet 
limitations as well as to optimize the beam time allocation.
The kinematic 
settings are shown in Table~\ref{tab:kine} along with the observed 
electron rate $R_e$ and the pion to electron ratio $R_{\pi^-}/R_e$ in the HRS. 
The highest electron rate occurred at RES II at approximately 600~kHz.

\begin{table}[!htp]
 \begin{center}
 \begin{tabular}{c|c|c|c|c|c|c}\hline\hline
  Kine\#  & HRS           & $E_b$ (GeV) & $\theta_0$ & $E^\prime_0$ (GeV)
  & $R_e$(kHz) & $R_{\pi^-}/R_e$ \\\hline
  DIS\#1  & Left          & 6.067 & $12.9^\circ$ & $3.66$ &$\approx 210$& $\approx 0.5$\\
  DIS\#2  & Left \& Right & 6.067 & $20.0^\circ$ & $2.63$ &$\approx 18$ & $\approx 3.3$\\
  RES I   & Left          & 4.867& $12.9^\circ$ & $4.0$   &$\approx 300$ & $<\approx 0.25$\\
  RES II  & Left          & 4.867 & $12.9^\circ$ & $3.55$ &$\approx 600$ & $<\approx 0.25$\\
  RES III & Right         & 4.867 & $12.9^\circ$ & $3.1$ & $\approx 400$ & $<\approx 0.4$\\
  RES IV  & Left          & 6.067  & $15^\circ$   & $3.66$ & $\approx 80$  & $<\approx 0.6 $\\
  RES V   & Left          & 6.067  & $14^\circ$   & $3.66$ & $\approx 130$ & $<\approx 0.7$\\
 \hline\hline
 \end{tabular}
 \caption{Overview of kinematics settings during the experiment, including: 
the beam energy $E_b$, the spectrometer central angle setting $\theta_0$ and central
momentum setting $E^\prime_0$, the observed electron rate $R_e$ and the 
$\pi^-/e$ ratio $R_{\pi^-}/R_e$.}\label{tab:kine}
 \end{center}
\end{table}

\section{DAQ PID Performance}\label{sec:pid}
\begin{figure}[!ht]
\includegraphics[width=\textwidth]{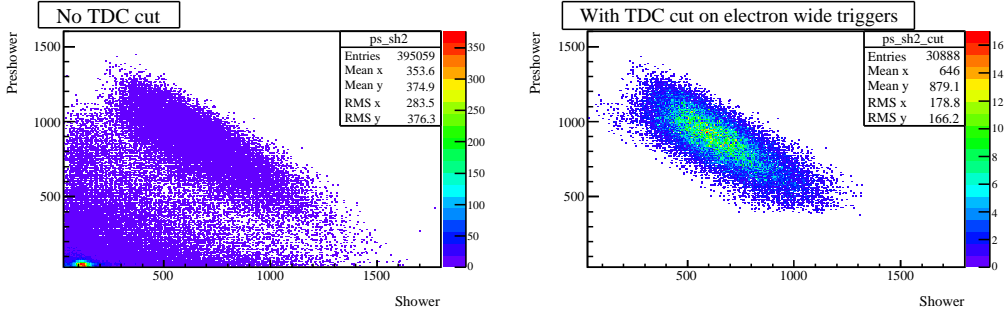}
\hspace*{-0.2cm}
\caption{Preshower vs. Shower ADC data (sum of 8 blocks each) for 
group 2 on the Left HRS,
without the fbTDC cut (left panel) and with cut on the group 2 electron wide trigger fbTDC
signal (right panel). This clearly shows the thresholds of the preshower and the total
shower signals, indicating that the DAQ is selecting the correct events
as electrons. %The cuts can be adjusted by changing the discriminator thresholds.
%%The events near the vertical axis, around ADC channels (200,1000), 
%%are electrons that deposited energy in overlapping blocks between group 2 and group 1 
%%(or group 3) and are recorded by the other group.
}
\label{fig:showerspectrum}
\end{figure}

The PID performance of the DAQ system was studied with calibration runs taken at low beam currents using 
fbTDC signals along with ADC data of all detector signals recorded by the standard DAQ. 
Events that triggered the DAQ would appear as a timing peak in the corresponding fbTDC 
spectrum of the standard DAQ, and a cut on this peak can be used to select those events. 
Figure~\ref{fig:showerspectrum} shows the preshower vs. shower
signals for group 2 on the Left HRS. A comparison between no fbTDC cut and with 
cut on the fbTDC signal of the electron wide trigger from this group clearly shows the 
hardware PID cuts.

%%%%%%%%%%%%%%%%
\begin{figure}[!ht]
\includegraphics[width=0.5\textwidth,angle=0]{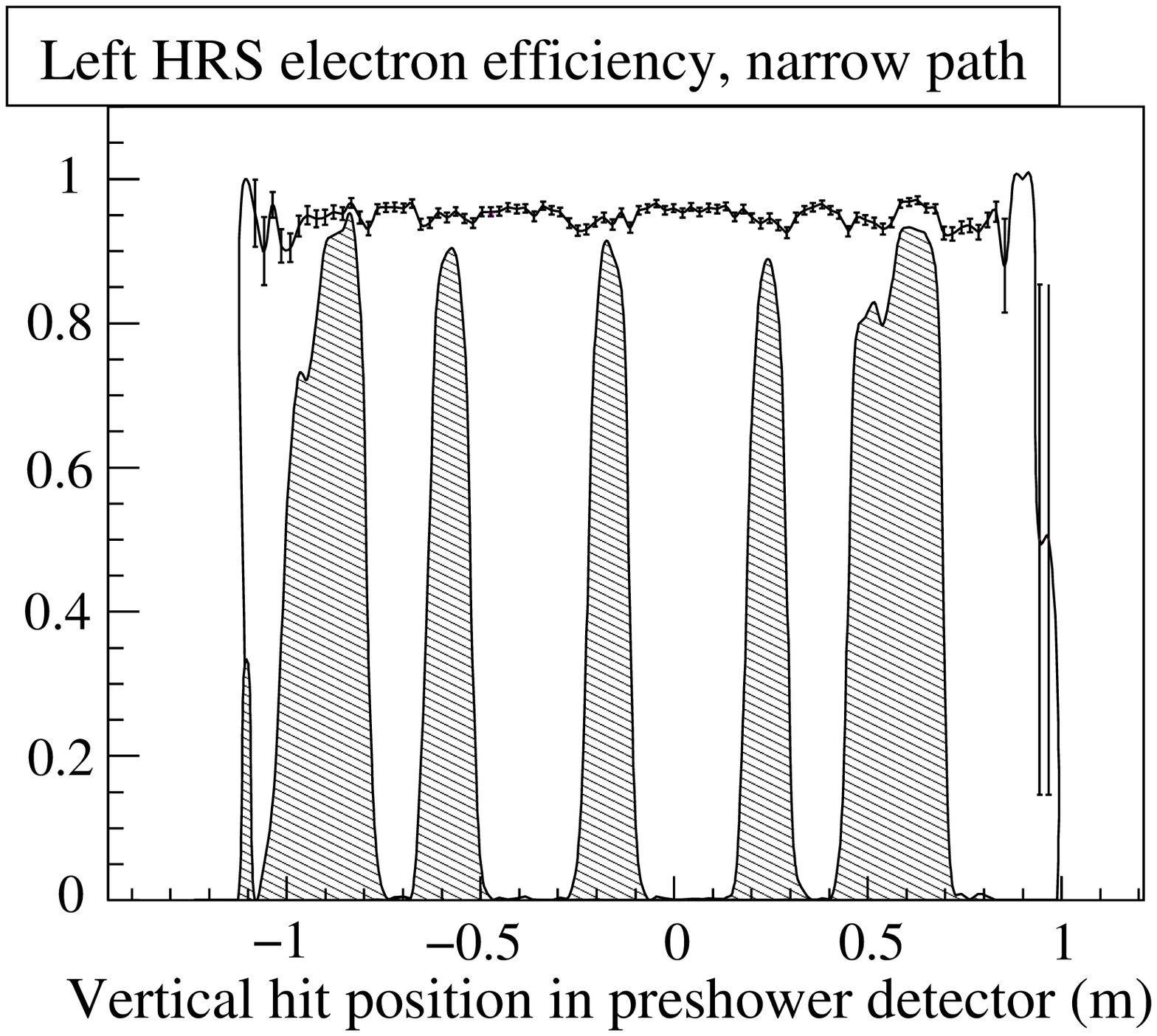}
\includegraphics[width=0.5\textwidth,angle=0]{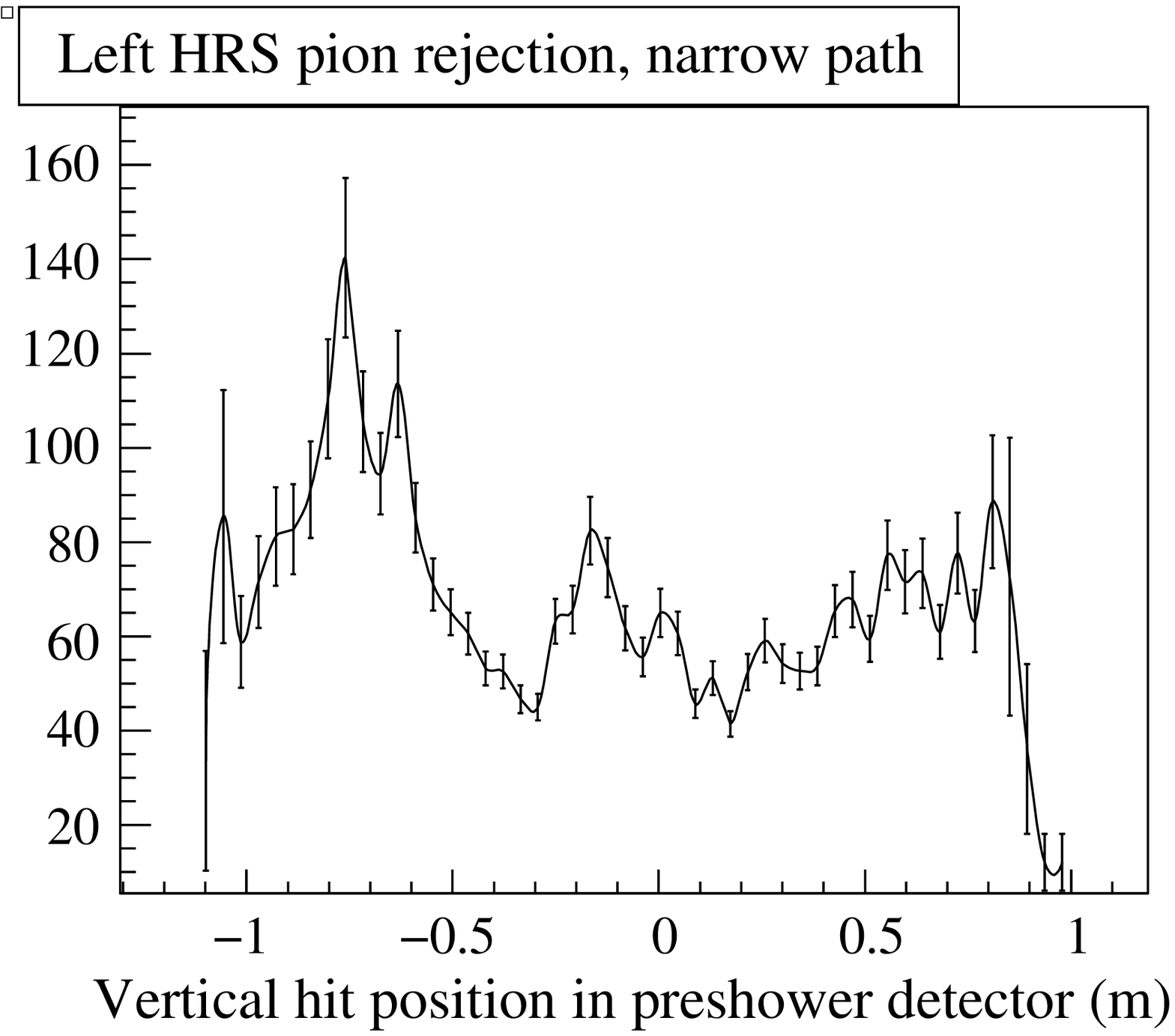}
\caption{Electron detection efficiency (left) and pion rejection factor (right) 
vs. vertical (dispersive) hit position of the particle in the preshower detector 
for the narrow electron triggers in the Left HRS. An 8-minute run with a reduced beam
current of $2$~$\mu$A at kinematics DIS \#2 was used in this evaluation. %run 26252
For electron efficiencies, the total efficiency and the statistical error bars are shown as the curve, while the 
shaded area indicates events that were recorded by two adjacent groups. 
The average electron efficiency achieved by the lead glass detector alone 
for this run is $[94.60\pm 0.11$(stat.)$]\%$ and the average
pion rejection factor is $[76.2\pm 1.5$(stat.)$]:1$.
PID performance for the wide path and the Right HRS are similar.
}
\label{fig:pidLeft}
\end{figure}

Electron efficiency and pion rejection factors of the lead-glass detector on 
the Left HRS during a one-hour run are 
shown in Fig.~\ref{fig:pidLeft} as functions of the location of the hit
of the particle in the preshower detector. PID performance on the Right HRS
is similar.
Electron efficiency from wide groups is slightly higher than from narrow groups
because there is less event loss due to timing misalignment when taking the
coincidence between the preshower and the total shower discriminator outputs.
Variations in the electron efficiency across the spectrometer acceptance 
effectively influence the $Q^2$ of the measurement. For this reason, 
low-rate calibration 
data were taken daily during the experiment to monitor the DAQ PID performance, 
and corrections were applied to the asymmetry data. 

\begin{figure}[!ht]
\begin{center}
%color version
\includegraphics[width=0.6\textwidth]{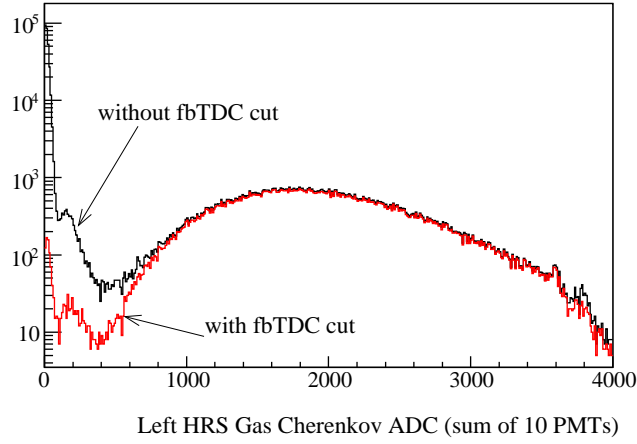}
%black&white version
%\includegraphics[width=0.6\textwidth]{Cherenkov_edit_bw.eps}
\end{center}
\caption{[{\it Color online}] Gas Cherenkov ADC data (sum of 10 PMTs)
for the Left HRS during a one-hour run at kinematics DIS \#2, 
with a fbTDC cut on the Cherenkov discriminator output 
and without. The beam current during this run was about 100~$\mu$A,
the incident electron rate on the detector was about 23~kHz with a pion
to electron rate ratio of approximately 3.5. The electron efficiency
achieved by the gas Cherenkov alone 
for this kinematics on the Left HRS was approximately 99\% 
with a pion rejection of approximately 300:1, see 
Table~\ref{tab:pidperformance_dis}.
The discriminator clearly selected electrons while
rejecting pions.  
}
\label{fig:cerspectrum}
\end{figure}

The gas Cherenkov detector signals were read out by 10 PMTs on both the Left and the Right HRS.
Signals from all 10 PMTs were summed in an analog-sum module and sent to a discriminator. 
The discriminator output was sent to the DAQ (as shown in Fig.~\ref{fig:daqflowchart})
as well as to fbTDCs. Figure~\ref{fig:cerspectrum} shows the Cherenkov ADC sum
with and without the fbTDC cut, which clearly shows the capability of rejecting
pions. 

As described in the Introduction, pion contamination in the electron trigger 
would affect the measured electron asymmetry as $A_m = (1-f_{\pi/e})A_e + f_{\pi/e} A_\pi$
where $A_m$ and $A_e$ are the measured and the true electron asymmetries,
respectively, 
and $A_\pi$ is the parity violation asymmetry of pion production.
The pion contamination in the electron trigger, $f_{\pi/e}$, comes from two effects:
There is a small possibility that a pion could trigger both the lead-glass 
and the gas Cherenkov detectors, causing a false electron trigger output. This possibility 
is determined by the direct combination of the pion rejection factors of the two detectors and is 
below $10^{-4}$. A larger effect comes from the width of the electron GATE signal: Since 
each coincidence between the gas Cherenkov and the scintillator signals would open
the electron counting GATE by 150~ns, while the DAQ deadtime of the
lead-glass detector is less than this value, pions that arrived after the DAQ deadtime
but before the closing of the electron GATE signal would cause a false electron trigger. The 
sum of the two effects can be written as
\begin{eqnarray}
 f_{\pi/e,n(w)} &=& \frac{R_\pi\eta^{GC}_\pi\eta^{LG}_\pi}{R_e\eta_e^{GC}\eta_e^{LG}}
   +\frac{R_\pi\eta^{LG}_\pi\{R_e\eta_e^{GC}[150~\mathrm{ns}-\tau_{n(w)}]\}}
  {R_e\eta_e^{GC}\eta_e^{LG}}
\end{eqnarray}
where $R_e$ and $R_\pi$ are the input electron and the pion rates, respectively; 
$\eta_e^{LG (GC)}$ is the electron detection efficiency of the lead-glass (gas Cherenkov) 
detectors, and $\eta_\pi^{LG (GC)}$ is the pion detection efficiency, i.e., the inverse 
of the rejection factor, of the lead-glass (gas Cherenkov) detector.
The DAQ group deadtime of the lead-glass detector for the narrow (wide) path, $\tau_{n(w)}$,
is approximately 60~ns (100-110 ns) and the analysis obtaining these results 
will be presented in the next section. The term $R_e\eta_e^{GC}[150~\mathrm{ns}-\tau_{n(w)}]$
gives the probability of a pion's arriving within a valid electron GATE signal and thus such a pion can not be rejected by the gas Cherenkov detector.

The electron detection efficiency and pion rejection factor averaged throughout
the data production period are shown in Tables~\ref{tab:pidperformance_dis} and~\ref{tab:pidperformance_res} for DIS and resonance kinematics, respectively, along with the resulting pion contamination $f_{\pi/e}$ evaluated 
separately for the narrow and the wide paths.
\begin{table}[!htp]
 \begin{center}
  \begin{tabular}{|c|c|c|c|}
\hline\hline
    \multicolumn{4}{|c|} {DIS Kinematics and Spectrometer combinations}\\\hline
            &  DIS\# 1                &  \multicolumn{2}{c|}{DIS\# 2} \\\hline
 HRS        &  Left                   & Left                   & Right  \\\hline
    \multicolumn{4}{|c|} {Electron detection efficiency $\eta_e$ ($\%$)}\\\hline
 GC         &  $99.14\pm 0.02$    & $99.03\pm 0.03$    & $98.19\pm 0.06$ \\
 LG, n      &  $91.93\pm 0.04$    & $94.50\pm 0.06$    & $94.36\pm 0.04$ \\
 LG, w      &  $92.88\pm 0.04$    & $95.79\pm 0.06$    & $95.23\pm 0.04$ \\
 GC+LG, n   & $91.14\pm 0.04$     & $93.58\pm 0.06$    & $92.65\pm 0.07$ \\
 GC+LG, w   & $92.08\pm 0.04$     & $94.86\pm 0.06$    & $93.51\pm 0.07$ \\\hline
    \multicolumn{4}{|c|} {Pion rejection $1/\eta_\pi$}\\\hline
 GC         &  $158.6\pm 3.5$    & $301.2\pm 5.2$      & $414.3\pm 6.2$\\
 LG, n      &  $101.5\pm 1.6$    & $78.9\pm 0.9$       & $72.7\pm 0.3$ \\
 LG, w      &  $103.9\pm 1.7$    & $81.5\pm 1.0$       & $74.3\pm 0.3$ \\\hline
    \multicolumn{4}{|c|} {Pion contamination in the electron trigger $f_{\pi/e}$, narrow path ($\times 10^{-4}$)}\\\hline
% $R_\pi/R_e$   & $\approx 0.5$           & $\approx 3.3$           & $\approx 3.3$ \\
 $f_{\pi/e,n}$ & $1.07$    & $1.97$     & $1.30$ \\
 %$\Delta f_{\pi/e,n}$
 (stat.)     & $\pm 0.02$ & $\pm 0.03$ & $\pm 0.01$ \\
 (syst.)     & $\pm 0.24$ & $\pm 0.18$ & $\pm 0.10$ \\
 (total)     & $\pm 0.24$ & $\pm 0.18$ & $\pm 0.10$ \\\hline
    \multicolumn{4}{|c|} {Pion contamination in the electron trigger $f_{\pi/e}$, wide path ($\times 10^{-4}$)}\\\hline
 $f_{\pi/e,w}$ & $0.72$     & $1.64$     & $0.92$ \\
 %$\Delta f_{\pi/e,w}$
 (stat.)     & $\pm 0.01$ & $\pm 0.03$ & $\pm 0.01$ \\
 (syst.)     & $\pm 0.22$ & $\pm 0.17$ & $\pm 0.13$ \\
 (total)     & $\pm 0.22$ & $\pm 0.17$ & $\pm 0.13$ \\
   \hline\hline
  \end{tabular}
 \end{center}
\vspace*{0.3cm}
 \caption{Average electron detection efficiency and pion rejection factor 
of electron triggers achieved for the DIS kinematics 
through the lead glass (LG) and the gas Cherenkov (GC) detectors, 
respectively, and the combined performance. The error bars of the efficiencies and the 
rejection factors are statistical only. The error bars for $f_{\pi/e}$, 
$\Delta f_{\pi/e,n(w)}$, are shown separately
for statistical uncertainties, systematic uncertainties due to our understanding of the 
rates, detector efficiencies and deadtimes, and day-to-day variations in the measured
detector efficiencies.
%while the error on $f_{\pi/e}$ includes both statistical and systematic uncertainties.
}\label{tab:pidperformance_dis}
\end{table}

\begin{table}[!htp]
 \begin{center}
  \begin{tabular}{|c|c|c|c|c|c|}
\hline\hline
    \multicolumn{6}{|c|} {Resonance Kinematics and Spectrometer combinations}\\\hline
             &  RES I                  & RES II & RES III & RES IV  & RES V   \\\hline
 HRS         &  Left                   & Left   & Right & Left & Left \\ \hline
    \multicolumn{6}{|c|} {Electron detection efficiency $\eta_e$ ($\%$)}\\\hline
 GC          &  $99.16\pm 0.09$    & $99.17\pm 0.13$    & $98.59\pm 0.11$     & $99.41\pm 0.07$    & $99.16\pm 0.11$ \\
 LG, n       &  $97.73\pm 0.07$    & $97.13\pm 0.07$    & $98.14\pm 0.06$     & $84.71\pm 0.18$    & $84.66\pm 0.21$ \\
 LG, w       &  $98.32\pm 0.07$    & $97.83\pm 0.08$    & $98.56\pm 0.06$     & $85.31\pm 0.18$    & $85.88\pm 0.23$ \\
 GC+LG, n    & $96.91\pm 0.11$     & $96.32\pm 0.15$    & $96.76\pm 0.12$     & $84.20\pm 0.20$    & $83.95\pm 0.24$ \\
 GC+LG, w    & $97.49\pm 0.11$     & $97.02\pm 0.15$    & $97.17\pm 0.13$    & $84.80\pm 0.20$    & $85.16\pm 0.26$  \\\hline
    \multicolumn{6}{|c|} {Pion rejection $1/\eta_\pi$}\\\hline
 GC          &  $82.8\pm 9.2$    & $97.7\pm 10.5$      & $195.0\pm 24.5$    & $149.6\pm 10.2$      & $151.4\pm 11.5$\\
 LG, n       &  $43.6\pm 4.0$    & $57.4\pm 5.4$       & $37.0\pm 0.9$     & $182.4\pm 15.1$      & $207.2\pm 20.5$\\
 LG, w       &  $39.4\pm 3.6$    & $53.5\pm 5.1$       & $33.9\pm 0.9$     & $171.4\pm 14.1$      & $201.1\pm 23.5$\\\hline
    \multicolumn{6}{|c|} {Pion contamination in the electron trigger $f_{\pi/e}$, narrow path ($\times 10^{-4}$)}\\\hline
% $R_\pi/R_e$   & $<\approx 0.25$           & $<\approx 0.25$           & $<\approx 0.4$            & $<\approx 0.6$           & $<\approx 0.7$ \\
 $f_{\pi/e,n}$ & $0.79$    & $2.40$     & $3.82$    & $0.26$     & $0.45$ \\
 %$\Delta f_{\pi/e,n}$
 (stat.)     & $\pm 0.02$ & $\pm 0.06$ & $\pm 0.02$ & $\pm 0.01$ & $\pm 0.01$ \\
 (syst.)     & $\pm 0.11$ & $\pm 0.19$ & $\pm 0.23$ & $\pm 0.02$ & $\pm 0.03$ \\
 (total)     & $\pm 0.11$ & $\pm 0.20$ & $\pm 0.23$ & $\pm 0.03$ & $\pm 0.03$ \\\hline
    \multicolumn{6}{|c|} {Pion contamination in the electron trigger $f_{\pi/e}$, wide path ($\times 10^{-4}$)}\\\hline
 $f_{\pi/e,w}$ & $0.54$     & $1.50$     & $2.14$    & $0.22$     & $0.32$ \\
 %$\Delta f_{\pi/e,w}$
 (stat.)     & $\pm 0.02$ & $\pm 0.04$ & $\pm 0.02$ & $\pm 0.01$ & $\pm 0.01$ \\
 (syst.)     & $\pm 0.14$ & $\pm 0.25$ & $\pm 0.48$ & $\pm 0.03$ & $\pm 0.04$ \\
 (total)     & $\pm 0.15$ & $\pm 0.25$ & $\pm 0.48$ & $\pm 0.03$ & $\pm 0.04$ \\
   \hline\hline
  \end{tabular}
 \end{center}
\vspace*{0.3cm}
 \caption{Average electron detection efficiency and pion rejection factor 
of electron triggers achieved for the resonance kinematics 
through the lead glass (LG) and the gas Cherenkov (GC) detectors, 
respectively, and the combined performance. The error bars of the efficiencies and the 
rejection factors are statistical only. The error bars for $f_{\pi/e}$, 
$\Delta f_{\pi/e,n(w)}$, are shown separately
for statistical uncertainties, systematic uncertainties due to our understanding of the 
rates, detector efficiencies and deadtimes, and day-to-day variations in the measured
detector efficiencies.
%while the error on $f_{\pi/e}$ includes both statistical and systematic uncertainties.
}\label{tab:pidperformance_res}
\end{table}

As shown in Tables~\ref{tab:pidperformance_dis}-\ref{tab:pidperformance_res}, 
the overall pion contamination was
on the order of $2\times 10^{-4}$ or lower throughout the experiment. 
Because pions are produced from nucleon resonance decays, 
the parity violation asymmetry of pion production is expected to be 
no larger than that of scattered electrons with the same momentum.
This was confirmed by asymmetries formed from pion triggers during this experiment.
The uncertainty in the electron asymmetry 
due to pion contamination is therefore on the order of $2\times 10^{-4}$ and is
negligible compared with the $3-4\%$ statistical uncertainty.

To understand fully the effect of pion background on the measured electron asymmetry,
it is important to extract asymmetries of the pion background to confirm that 
they are indeed smaller than the electron asymmetry. A complete PID analysis was 
carried out on the pion triggers of the DAQ where  
the electron contamination in the pion trigger $f_{e/\pi}$ was evaluated in a similar method as 
$f_{\pi/e}$ above, following
\begin{equation}
 f_{e/\pi, n(w)} = \frac{R_e\xi^{GC}_e \xi^{LG}_e}{R_\pi\xi_\pi^{GC}\xi_\pi^{LG}}
   +\frac{R_e\xi^{LG}_e\{R_\pi\xi_\pi^{GC}[150~\mathrm{ns}-\tau_{n(w)}]\}}
  {R_\pi\xi_\pi^{GC}\xi_\pi^{LG}}
\end{equation}
where as before $R_e$ and $R_\pi$ are the electron and the pion rates incident on the detectors, 
respectively; 
the detection efficiencies $\xi$ are now defined for the pion triggers of the DAQ: 
$\xi_e^{LG (GC)}$ is the electron detection efficiency of the lead-glass (gas Cherenkov) 
detectors, and $\xi_\pi^{LG (GC)}$ is the pion detection efficiency of the lead-glass 
(gas Cherenkov) detector. 
Although the goal of the pion triggers is to collect pions, only the gas Cherenkov played a role
in rejecting electrons in the pion trigger, and all electrons would form valid pion triggers
in the lead-glass counters. Therefore $\xi_e^{LG}\approx 1$ 
and the electron contamination is high. Results for electron contamination in the pion trigger
are summarized in Tables~\ref{tab:pidperformance_pion_dis} and~\ref{tab:pidperformance_pion_res}.

\begin{table}[!htp]
 \begin{center}
  \begin{tabular}{|c|c|c|c|}
\hline\hline
    \multicolumn{4}{|c|} {Kinematics and Spectrometer Combinations}\\\hline
            & DIS\#1 & \multicolumn{2}{c|}{DIS\#2} \\\hline
 HRS        &  Left                 & Left   & Right  \\\hline
    \multicolumn{4}{|c|} {Pion detection efficiency $\xi_\pi$ ($\%$)}\\\hline
 GC         & $99.52\pm 0.01$   & $99.73\pm 0.01$    & $99.74\pm 0.01$ \\
 LG, n      & $21.67\pm 0.01$   & $79.72\pm 0.02$    & $15.61\pm 0.01$ \\
 LG, w      & $21.67\pm 0.01$   & $79.71\pm 0.02$    & $15.60\pm 0.01$ \\
 GC+LG, n   & $21.57\pm 0.01$   & $79.70\pm 0.02$    & $15.57\pm 0.01$ \\
 GC+LG, w   & $21.57\pm 0.01$   & $79.69\pm 0.02$    & $15.56\pm 0.01$ \\\hline
    \multicolumn{4}{|c|} {Electron rejection $1/\xi_e$}\\\hline
 GC         & $31.42\pm 0.78$    & $89.44\pm 2.48$    & $48.48\pm 1.55$\\
 LG, n      & $1.0468\pm 0.0003$ & $1.0487\pm 0.0005$ & $1.0271\pm 0.0002$ \\
 LG, w      & $1.0469\pm 0.0003$ & $1.0499\pm 0.0005$ & $1.0279\pm 0.0002$ \\\hline
    \multicolumn{4}{|c|} {Electron contamination in pion triggers $f_{e/\pi}$, narrow path}\\\hline
% actual rate $R_\pi/R_e$    & $\approx 0.5$         & $\approx 3.3$   & $\approx 3.3$ \\
 $f_{e/\pi,n}$& $0.2653$           & $0.0331$           & $0.0103$ \\
 %$\Delta f_{e/\pi,n}$
 (stat.)    & $\pm 0.0029$       & $\pm 0.0006$       & $\pm 0.0002$ \\
 (syst.)    & $\pm 0.0602$       & $\pm 0.0033$       & $\pm 0.0013$ \\
 (total)    & $\pm 0.0603$       & $\pm 0.0034$       & $\pm 0.0013$ \\\hline
    \multicolumn{4}{|c|} {Electron contamination in pion triggers $f_{e/\pi}$, wide path}\\\hline
 $f_{e/\pi,w}$ & $0.2176$           & $0.0281$           & $0.0091$ \\
 %$\Delta f_{e/\pi,w}$
 (stat.)    & $\pm 0.0029$       & $\pm 0.0006$       & $\pm 0.0002$ \\
 (syst.)    & $\pm 0.0573$       & $\pm 0.0036$       & $\pm 0.0012$ \\
 (total)    & $\pm 0.0573$       & $\pm 0.0037$       & $\pm 0.0013$ \\
% combined &  $[(1.61\pm 0.04)\times 10^4]:1$  & $[(2.38\pm 0.05)\times 10^4]:1$ & $[(3.01\pm 0.05)\times 10^4]:1$\\
   \hline\hline
  \end{tabular}
 \end{center}
\vspace*{0.3cm}
 \caption{Average pion detection efficiency and electron rejection factor of pion triggers
achieved for DIS kinematics through the lead glass (LG) and the gas Cherenkov (GC) detectors, 
respectively, and the combined performance. The error bars of the efficiencies and the 
rejection factors are statistical only. The error bars for $f_{e/\pi}$, 
$\Delta f_{e/\pi,n(w)}$, are shown separately
for statistical uncertainties, systematic uncertainties, and day-to-day variations in the measured
detector efficiencies.
%while the error on $f_{e/\pi}$ includes both statistical and systematic uncertainties.
\medskip
}\label{tab:pidperformance_pion_dis}
\end{table}

\begin{table}[!htp]
 \begin{center}
  \begin{tabular}{|c|c|c|c|c|c|}
\hline\hline
    \multicolumn{6}{|c|} {Kinematics and Spectrometer Combinations}\\\hline
             & RES I              & RES II             & RES III           & RES IV           & RES V \\ \hline
 HRS         &  Left              & Left              & Right              & Left             & Left  \\\hline
    \multicolumn{6}{|c|} {Pion detection efficiency $\xi_\pi$ ($\%$)}\\\hline
 GC          & $98.82\pm 0.13$    & $98.96\pm 0.11$    & $99.43\pm 0.07$   & $99.38\pm 0.04$  & $99.47\pm 0.04$ \\
 LG, n       & $26.27\pm 0.62$    & $25.65\pm 0.55$    & $82.78\pm 0.16$   & $21.16\pm 0.25$  & $20.69\pm 0.28$ \\
 LG, w       & $27.07\pm 0.65$    & $26.14\pm 0.57$    & $83.60\pm 0.17$   & $22.54\pm 0.26$  & $20.71\pm 0.28$ \\
 GC+LG, n    & $25.96\pm 0.64$    & $25.39\pm 0.56$    & $82.31\pm 0.17$   & $21.03\pm 0.25$  & $20.58\pm 0.28$ \\
 GC+LG, w    & $26.75\pm 0.66$    & $25.87\pm 0.58$    & $83.12\pm 0.18$   & $22.40\pm 0.26$  & $20.60\pm 0.28$ \\\hline
    \multicolumn{6}{|c|} {Electron rejection $1/\xi_e$}\\\hline
 GC          & $121.36$          & $118.33$            & $72.91$           & $101.43$         & $74.80$\\
             & $\pm 21.71$       & $\pm 34.02$         & $\pm 5.67$        & $\pm 16.59$      & $\pm 13.57$\\
 LG, n       & $1.0167$          & $1.0194$            & $1.0114$          & $1.0677$         & $1.0652$ \\
             & $\pm 0.0006$      & $\pm 0.0006$        & $\pm 0.0005$      & $\pm 0.0014$     & $\pm 0.0016$ \\
 LG, w       & $1.0167$          & $1.0105$            & $1.0064$          & $1.0344$         & $1.0541$ \\
             & $\pm 0.0006$      & $\pm 0.0006$        & $\pm 0.0005$      & $\pm 0.0013$     & $\pm 0.0016$ \\\hline
    \multicolumn{6}{|c|} {Electron contamination in pion triggers $f_{e/\pi}$, narrow path}\\\hline
% actual rate $R_\pi/R_e$    & $\approx 0.5$         & $\approx 3.3$   & $\approx 3.3$ \\
 $f_{e/\pi,n}$ & $0.4114$           & $0.3155$            & $0.0849$          & $0.1852$         & $0.1871$ \\
 %$\Delta f_{e/\pi,n}$
 (stat.)    & $\pm 0.0117$       & $\pm 0.0061$        & $\pm 0.0006$      & $\pm 0.0062$     & $\pm 0.0058$ \\
 (syst.)    & $\pm 0.0163$       & $\pm 0.0151$        & $\pm 0.0029$      & $\pm 0.0038$     & $\pm 0.0050$ \\
 (total)    & $\pm 0.0201$       & $\pm 0.0163$        & $\pm 0.0030$      & $\pm 0.0073$     & $\pm 0.0077$ \\\hline
    \multicolumn{6}{|c|} {Electron contamination in pion triggers $f_{e/\pi}$, wide path}\\\hline
 $f_{e/\pi,w}$& $0.3423$           & $0.2409$            & $0.0633$          & $0.1661$         & $0.1598$ \\
 %$\Delta f_{e/\pi,w}$
 (stat.)    & $\pm 0.0116$       & $\pm 0.0062$        & $\pm 0.0006$      & $\pm 0.0063$     & $\pm 0.0057$ \\
 (syst.)    & $\pm 0.0200$       & $\pm 0.0190$        & $\pm 0.0059$      & $\pm 0.0049$     & $\pm 0.0064$ \\
 (total)    & $\pm 0.0231$       & $\pm 0.0200$        & $\pm 0.0060$      & $\pm 0.0080$     & $\pm 0.0086$ \\
% combined &  $[(1.61\pm 0.04)\times 10^4]:1$  & $[(2.38\pm 0.05)\times 10^4]:1$ & $[(3.01\pm 0.05)\times 10^4]:1$\\
   \hline\hline
  \end{tabular}
 \end{center}
\vspace*{0.3cm}
 \caption{Average pion detection efficiency and electron rejection factor of pion triggers 
achieved for resonance kinematics
through the lead glass (LG) and the gas Cherenkov (GC) detectors, 
respectively, and the combined performance. The error bars of the efficiencies and the 
rejection factors are statistical only. The error bars for $f_{e/\pi}$, 
$\Delta f_{e/\pi,n(w)}$, are shown separately
for statistical uncertainties, systematic uncertainties, and day-to-day variations in the measured
detector efficiencies.
%while the error on $f_{e/\pi}$ includes both statistical and systematic uncertainties.
}\label{tab:pidperformance_pion_res}
\end{table}

\section{DAQ Deadtime}\label{sec:deadtime}
%
%%%%%%%%%%%%%%%%
Deadtime is the amount of time after an event during which the system is unable to record 
another event. Identifying the exact value of the deadtime is always a challenge 
in counting experiments. 
By having a narrow and a wide path, we can observe the trend in the deadtime: The wider
path should have higher deadtime.  By matching the observed trend with our simulation  
we can benchmark and confirm the result of our deadtime simulation.
In addition, dividing lead-glass blocks into groups greatly reduces the deadtime loss
in each group compared with summing all blocks together and forming only one final trigger.

To illustrate the importance of the deadtime, consider its effect on the asymmetry $A$.
For a simple system with only one contribution to the deadtime loss $\delta$, the observed
asymmetry $A_O$ is related to the true asymmetry $A$ according to $A_O = (1-\delta) A$.
\hskip 0.05in In this experiment $\delta$ was expected to be on the order of (1-2)\%. 
Since the statistical accuracy of the asymmetry is (3-4)\%, it was desirable to know $\delta$
with a (10-20)\% relative accuracy so that it would become a negligible systematic error.
The DAQ used in this experiment, however, was more complex and had three contributions to the deadtime 
as listed below:

\begin{enumerate}
 \item The ``group'' deadtime: deadtime due to discriminators and logical AND modules used to 
form group triggers.
 \item The ``GATE'' deadtime: deadtime from the GATE circuit that 
used scintillators and gas Cherenkov signals to form the GATE signals, which controlled 
the AND (OR) module of each group to form group electron (pion) triggers.
 \item The ``OR'' deadtime: deadtime due to the logical OR module used to combine all group 
triggers into final global triggers.
\end{enumerate}
The total deadtime is a combination of all three.  
In order to evaluate the DAQ deadtime, a full-scale trigger simulation is necessary. This 
trigger simulation will be described in the next section followed by results on the group, 
GATE, and OR deadtime as well as on the total deadtime correction that was applied to the 
asymmetry data. 

\subsection{Trigger Simulation}
\begin{figure}[!htp]
%color version
\includegraphics[width=0.5\textwidth,angle=0]{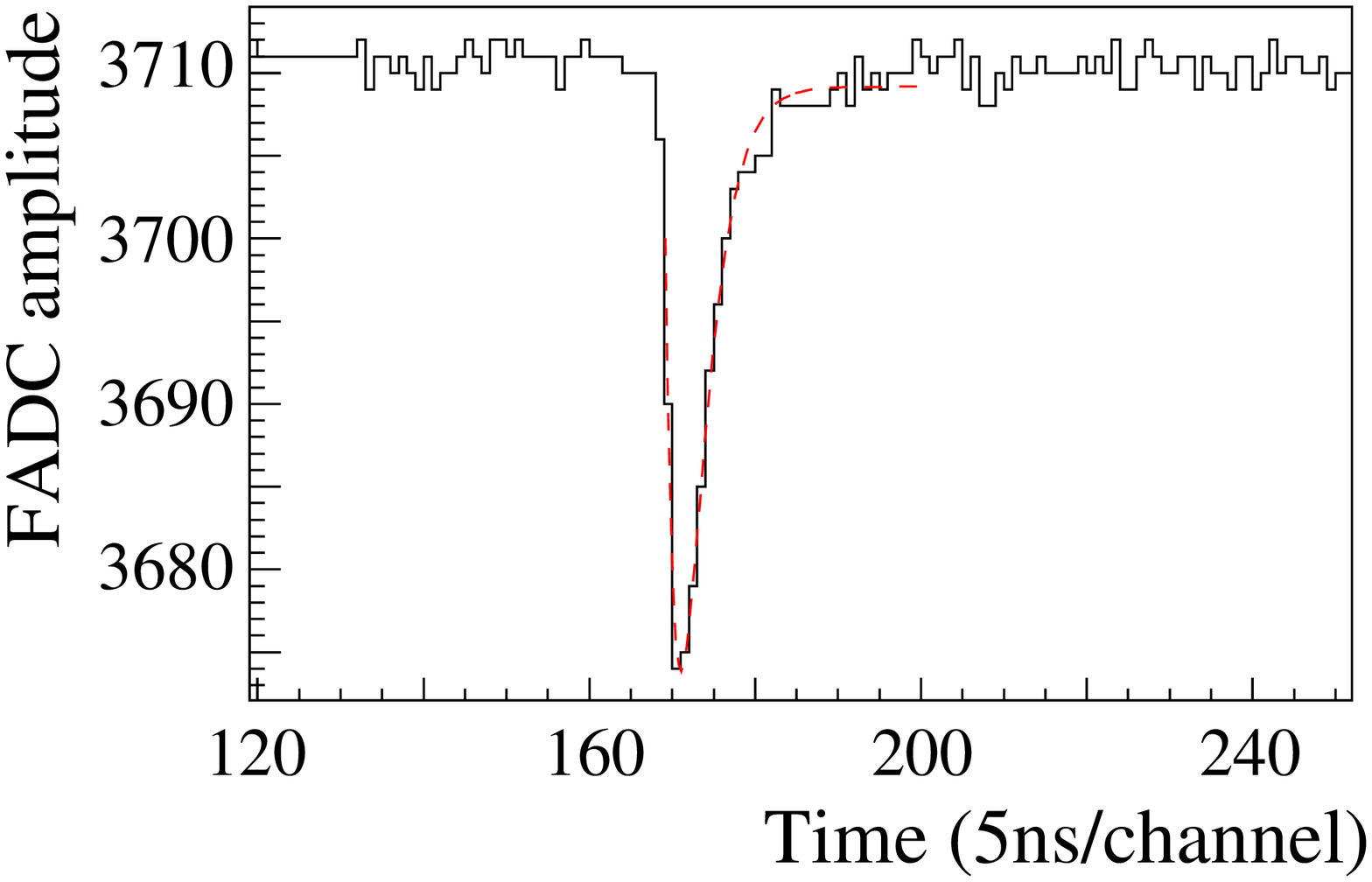}
\includegraphics[width=0.5\textwidth,angle=0]{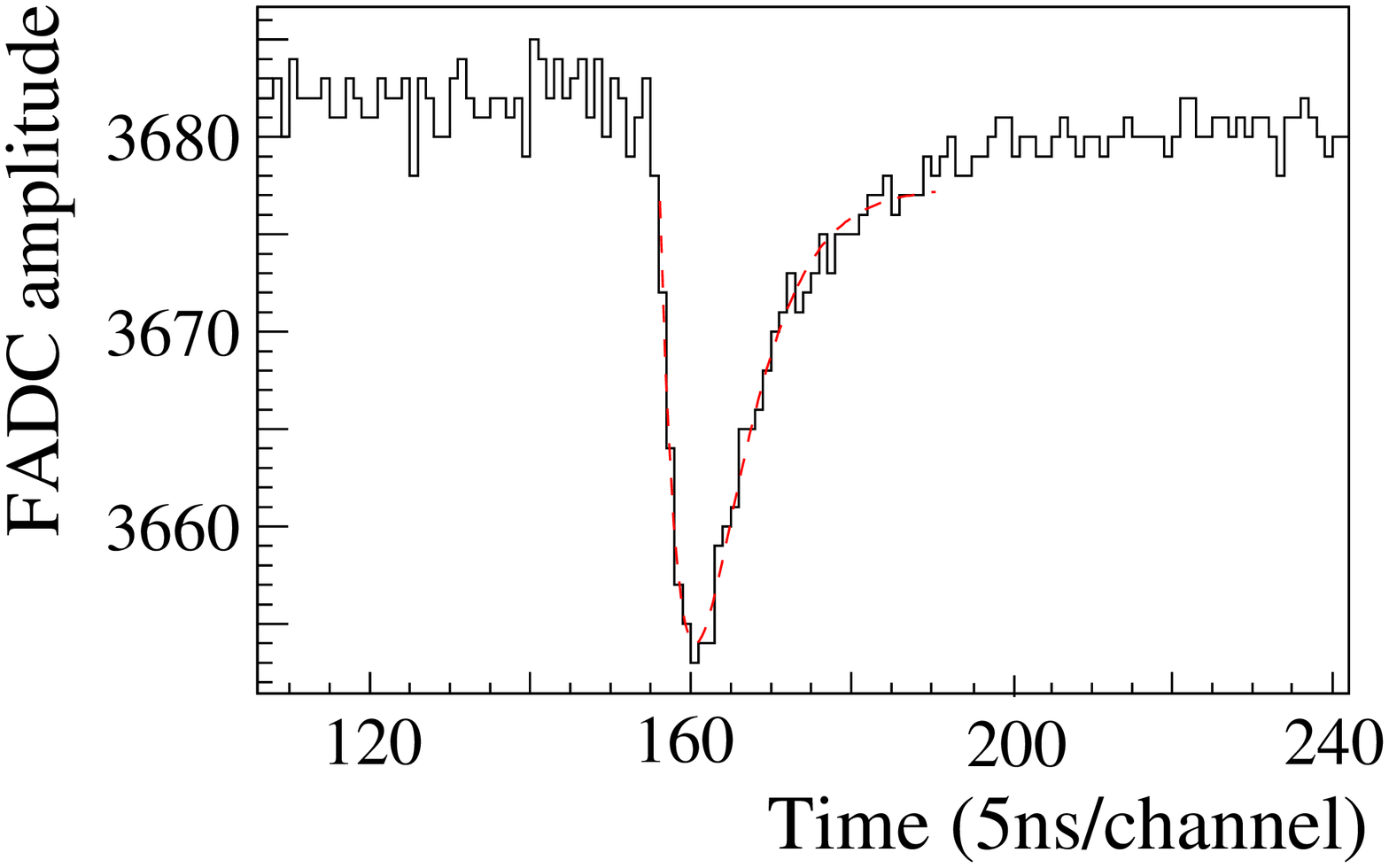}
%black&white version
%\includegraphics[width=0.5\textwidth,angle=0]{pscalib_edit_bw.eps}
%\includegraphics[width=0.5\textwidth,angle=0]{shcalib_edit_bw.eps}
\caption{[{\it Color online}] Calibration of time constants $\tau$ for Preshower 
(left) and Shower (right) of the Right HRS. The FADC snapshots (histograms) is compared 
with the fit $S(t)=A t e^{-t/\tau}$ (smooth dashed curves). The time constant $\tau$ was found to be approximately $11$~ns for the Right HRS Preshower, and $21-22$~ns for the Left HRS Preshower and Shower as well as for the Right HRS Shower.}\label{fig:hats_calib}
\end{figure}

The Hall A Trigger Simulation (HATS) was developed for the purpose of deadtime study for 
this experiment. The inputs to HATS include the analog signals for preshower, shower, 
scintillator and gas Cherenkov. The signal amplitudes were provided by ADC data from 
low-current runs, and the signal rates were from high-current production runs.
The rise and fall times for the preshower and shower SUM8 outputs play an 
important role in HATS. The signal shape is simulated by the function 
$S(t)=A t e^{-t/\tau}$, where $A$ is related to the amplitude of the signal, and the time 
constant $\tau$ was determined from FADC data, see Fig.~\ref{fig:hats_calib}.

\begin{figure}[!htp]
\centering
\includegraphics[width=1.0\textwidth]{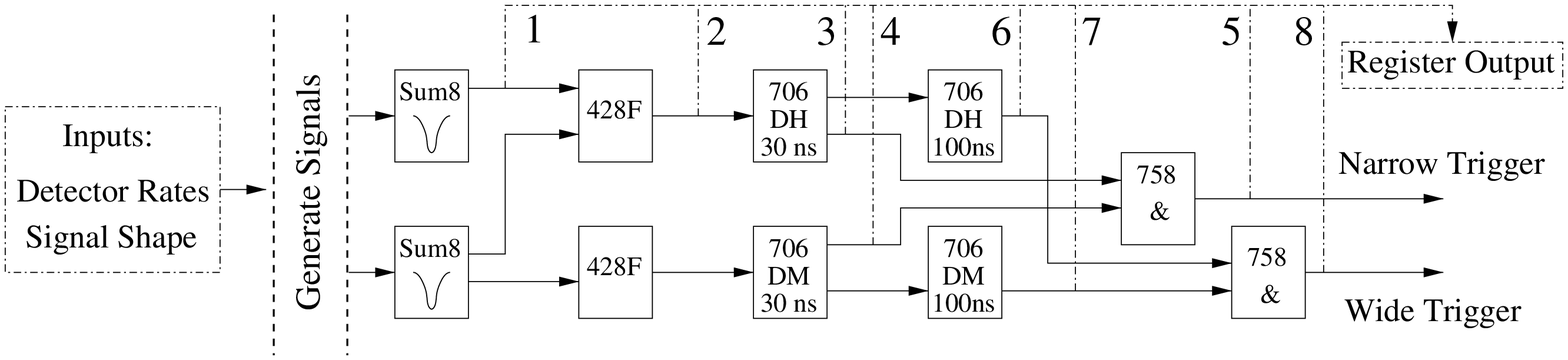}
\includegraphics[width=0.9\textwidth]{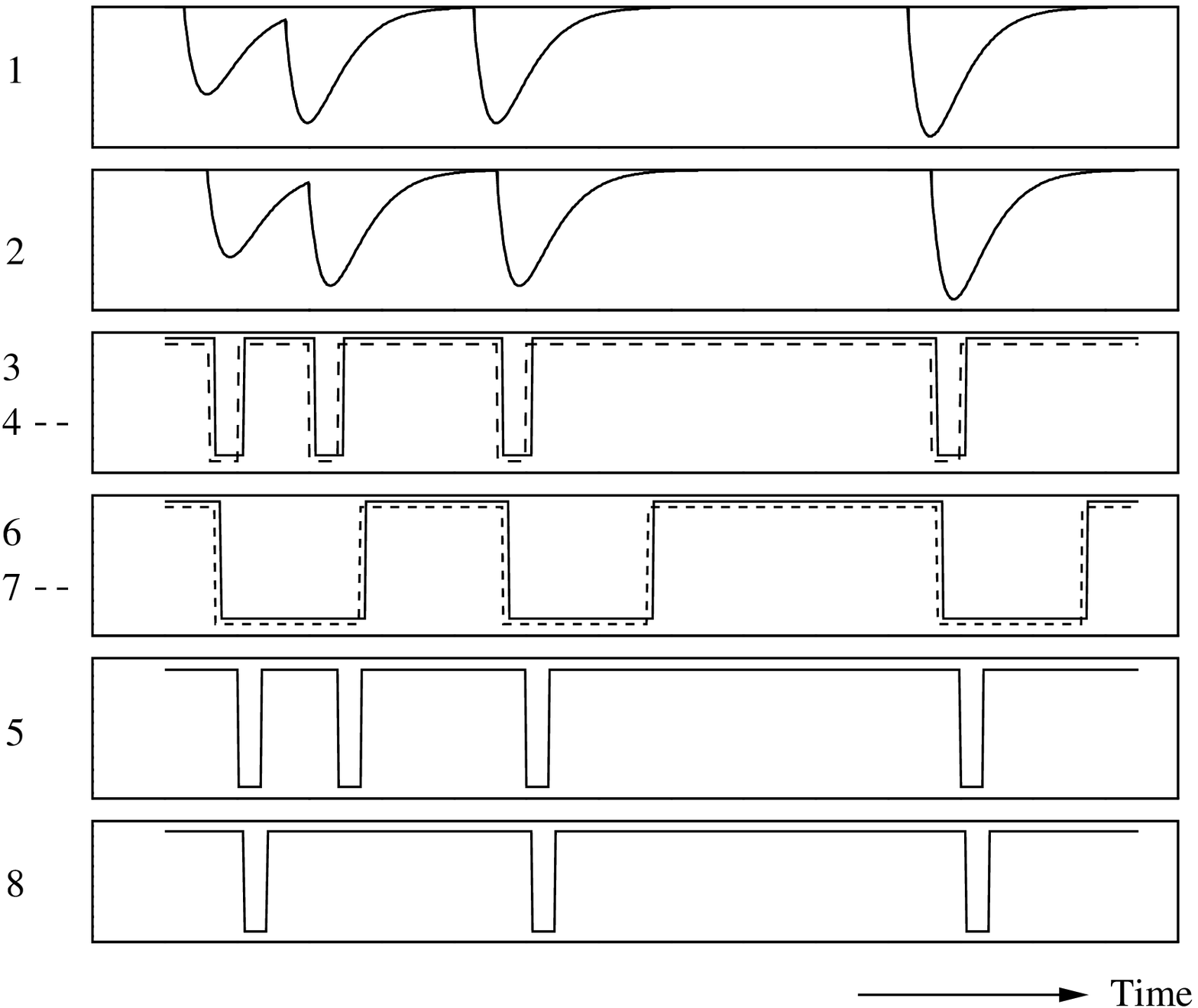}
\caption{ 
Top: A fraction of the group electron trigger. Each point corresponds to: 
1 -- Shower sum of the group; 2 -- Total shower sum of the group; 3 -- Total shower
discriminator output (high threshold), narrow path; 4 -- Preshower discriminator 
output (medium threshold), narrow path; 
5 -- group electron trigger, narrow path; 6 -- Total shower discriminator output, 
wide path; 7 -- Preshower discriminator output, wide path; 8 -- group electron
trigger, wide path. Bottom: Signals 1-8 as simulated by HATS. One can see that the second
physical event is recorded by the narrow path group trigger (5) but not the wide path
(8) due to deadtime loss. }\label{fig:hats_example}
\end{figure}

With the recorded DAQ electronics and delay cables, HATS first rebuilds the DAQ system
on the software level. At each nano-second, detector input signals are generated randomly 
according to the actual event rates and signal shape, and HATS
simulates output signals from all discriminators, AND, and OR modules. 
Figure ~\ref{fig:hats_example} shows a fraction of the DAQ electronics and the 
simulated results for a very short time period. By comparing output with input signals, 
HATS provides results on the fractional loss
due to deadtime for all group and global triggers with respect to the input signal.

\subsection{Group Deadtime Measurement}\label{sec:deadtime_group}
\begin{figure}[!ht]
\begin{center}
%color version
\includegraphics[width=0.8\textwidth]{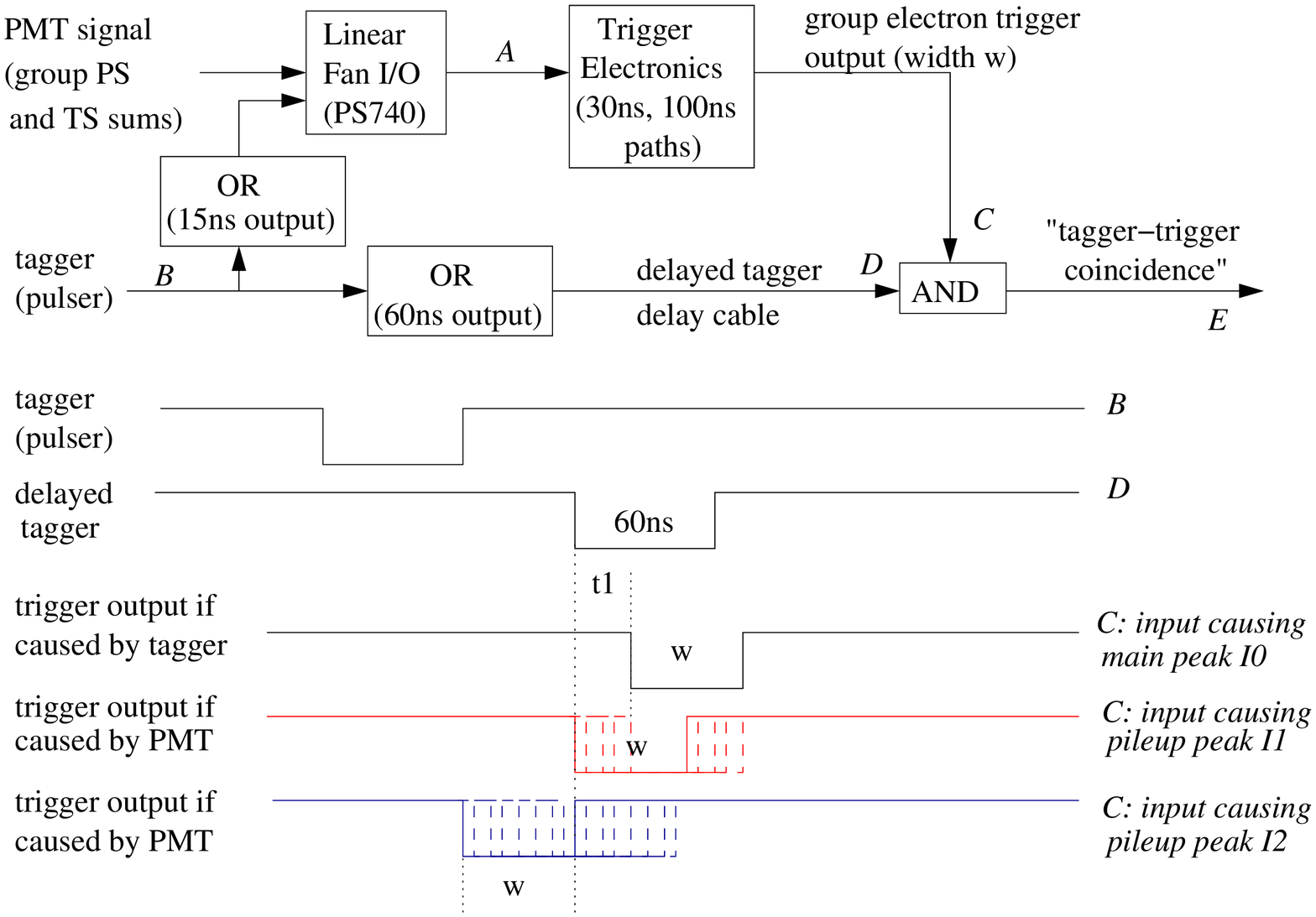}\\
%black&white version
%\includegraphics[width=0.8\textwidth]{Fig5_schematic_bw.eps}\\

\vspace*{0.5cm}
%color version
\includegraphics[width=0.5\textwidth]{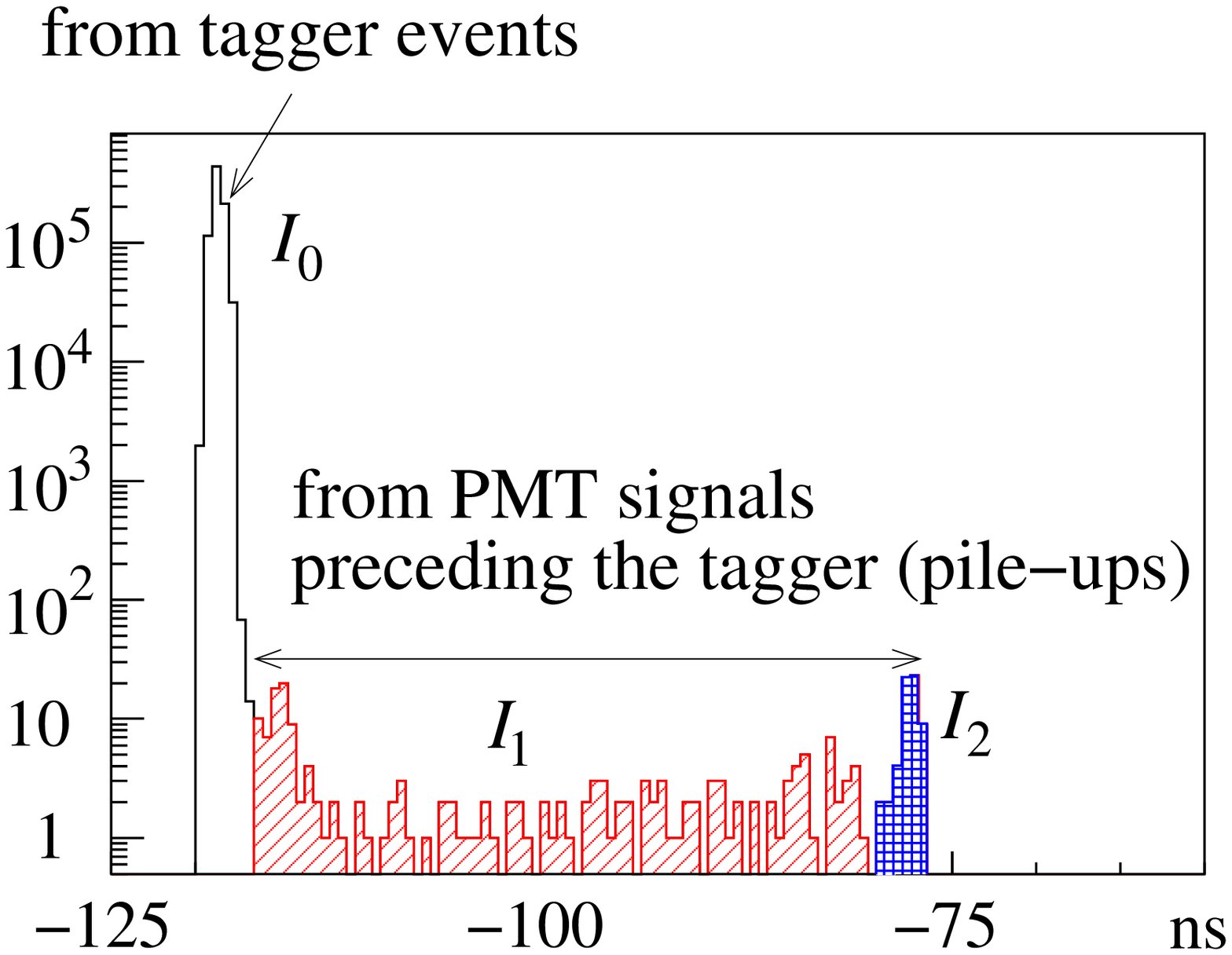}
%black&white version
%\includegraphics[width=0.5\textwidth]{pileup5583_edit_bw.eps}
\end{center}
\caption{[{\it Color online}] Top: schematic diagram for the tagger setup and signal timing sequence. The two
logical OR units immediately following the tagger input ``{\it B}'' serve as width adjusters.
Bottom: fbTDC spectrum for the relative timing between tagger-trigger coincidence
and the input tagger. The fbTDC module worked in a common stop and the multi-hit mode. 
Two different scenarios are 
shown: 1) Main peak $I_0$ (hollow peak): when there is no PMT signal preceding the tagger, the tagger triggers the 
DAQ and forms a tagger-trigger coincidence. 
2) Pileup events $I_1$ (light-shaded region) and $I_2$ (heavy-shaded region): when there is a PMT signal preceding the tagger by a time interval 
shorter than $w+t_1$, the PMT
signal triggers the DAQ and forms a tagger-trigger coincidence signal with the delayed tagger. 
}\label{fig:tagger}
\end{figure}

In order to study the group deadtime, a high rate pulser signal (``tagger'')
was mixed with the Cherenkov and all preshower and total shower signals 
using analog summing modules, see 
Figs.~\ref{fig:daqflowchart} and \ref{fig:tagger}.
In the absence of all detector signals, a tagger pulse produces without loss an electron
trigger output and a ``tagger-trigger coincidence'' pulse between this output and the 
``delayed tagger'' -- the tagger itself with an appropriate delay to account for the DAQ response time.
When high-rate detector signals are present, however, some of the tagger pulses 
would not be able
to trigger the DAQ due to deadtime. The deadtime loss in the electron trigger output
with respect to the tagger input has two components:
\begin{enumerate}
 \item The count loss $R_o/R_i$: When a detector PMT signal precedes the tagger signal by a time 
interval $\delta t$ shorter than the DAQ deadtime but longer than $w+t_1$, the 
tagger signal is lost and no coincidence output is formed. Here $w$ is the width of 
the electron trigger output and $t_1$ is the time interval by which the delayed tagger precedes the tagger's 
own trigger output, see Fig.~\ref{fig:tagger}. During the experiment $w$ was set to 15~ns 
for all groups, and $t_1$ was measured at the end of the experiment and found to be between 
20 and 40~ns for all narrow and wide groups of the two HRSs. 
 \item The pileup fraction $p$: When a PMT signal precedes the tagger signal by a time 
interval $\delta t$ shorter than $w+t_1$, there 
would be a coincidence output between the delayed tagger and the electron output triggered
by the detector PMT signal.
If furthermore $\delta t$ is less than the DAQ deadtime (which is possible for this experiment
since the deadtime is expected to be as long as 100~ns for the wide path), 
the tagger itself is lost due to deadtime, and
the tagger-trigger coincidence is a false count and should be subtracted. 
In the case where $\delta t$ is shorter than $w+t_1$ but longer than the DAQ deadtime 
(not possible for this experiment
but could happen in general), the tagger itself also triggers a
tagger-trigger coincidence, but in this case, there are two tagger-trigger coincidence 
events. Both are recorded by the fbTDC if working in the multi-hit mode, 
and one is a false count and should be subtracted. 

The pileup effect can be measured using the delay between the tagger-trigger coincidence output
and the input tagger.
This is illustrated in Fig.~\ref{fig:tagger} and the pileup effect contributes to both $I_1$ and $I_2$
regions of the fbTDC spectrum. 
The $I_1$ distribution is produced by PMT pulses that arrive after the delayed tagger signal 
but before the tagger signal would propagate through the trigger electronics. 
Peak $I_2$ occurs when a PMT pulse arrives at the coincidence module earlier than the delayed
tagger signal but which forms a coincidence with the delayed tagger signal, giving an output 
whose time is set by the latter.
Fractions of $I_1$ and $I_2$ relative to $I_0$ are expected to be
$I_1/I_0=R t_1$ and $I_2/I_0=Rw$, respectively, where $R$ is the PMT signal rate. 
The pileup effect was measured using fbTDC spectrum for electron narrow and wide
triggers for all groups. Data for $I_{1,2}$ extracted from fbTDC agree very well with the expected values.
\end{enumerate}

%%%%%%%%%%%%%%%%

\begin{figure}[!ht] 
 \begin{center}
%color version
  \includegraphics[width=0.6\textwidth,angle=0]{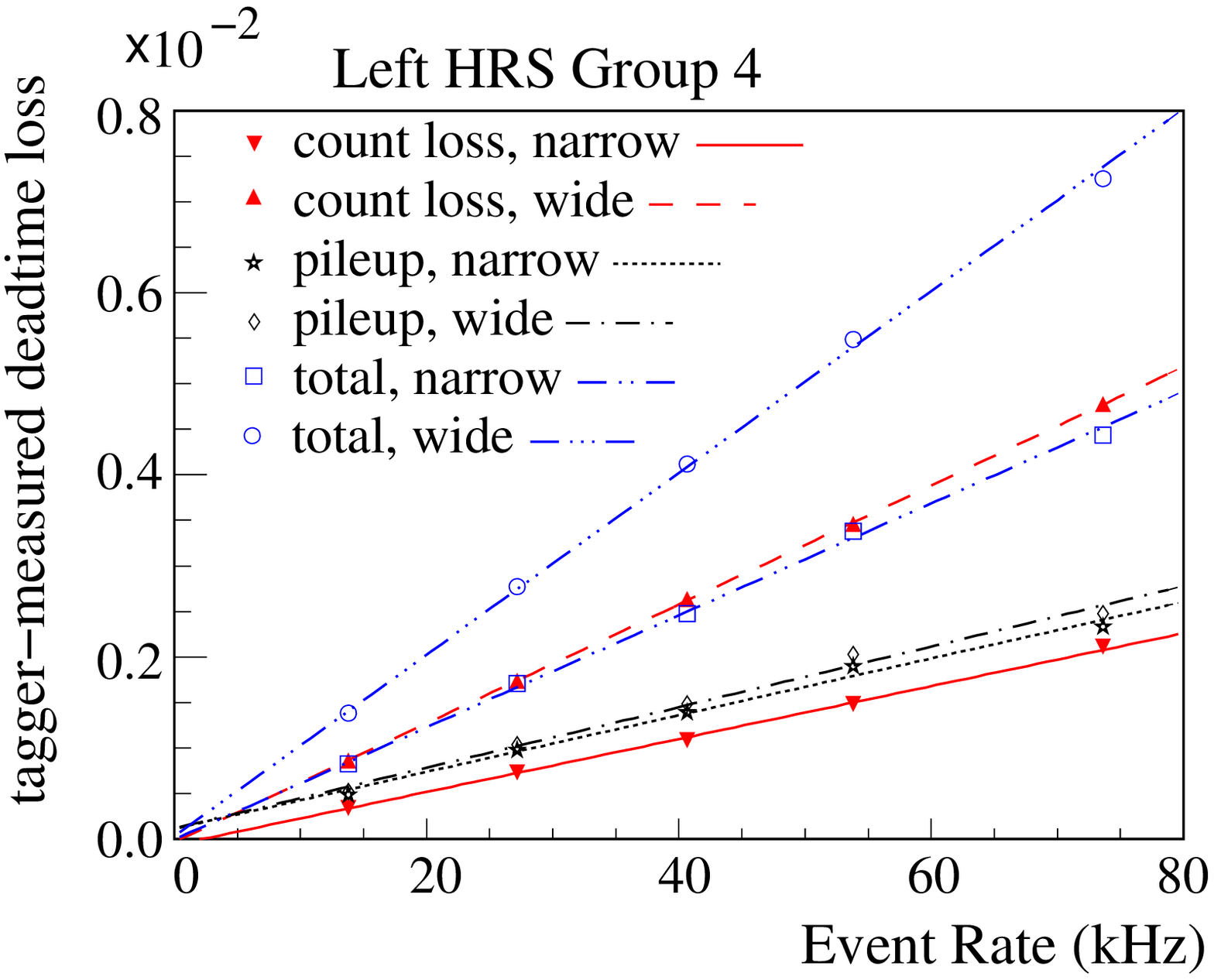}
  \includegraphics[width=0.6\textwidth,angle=0]{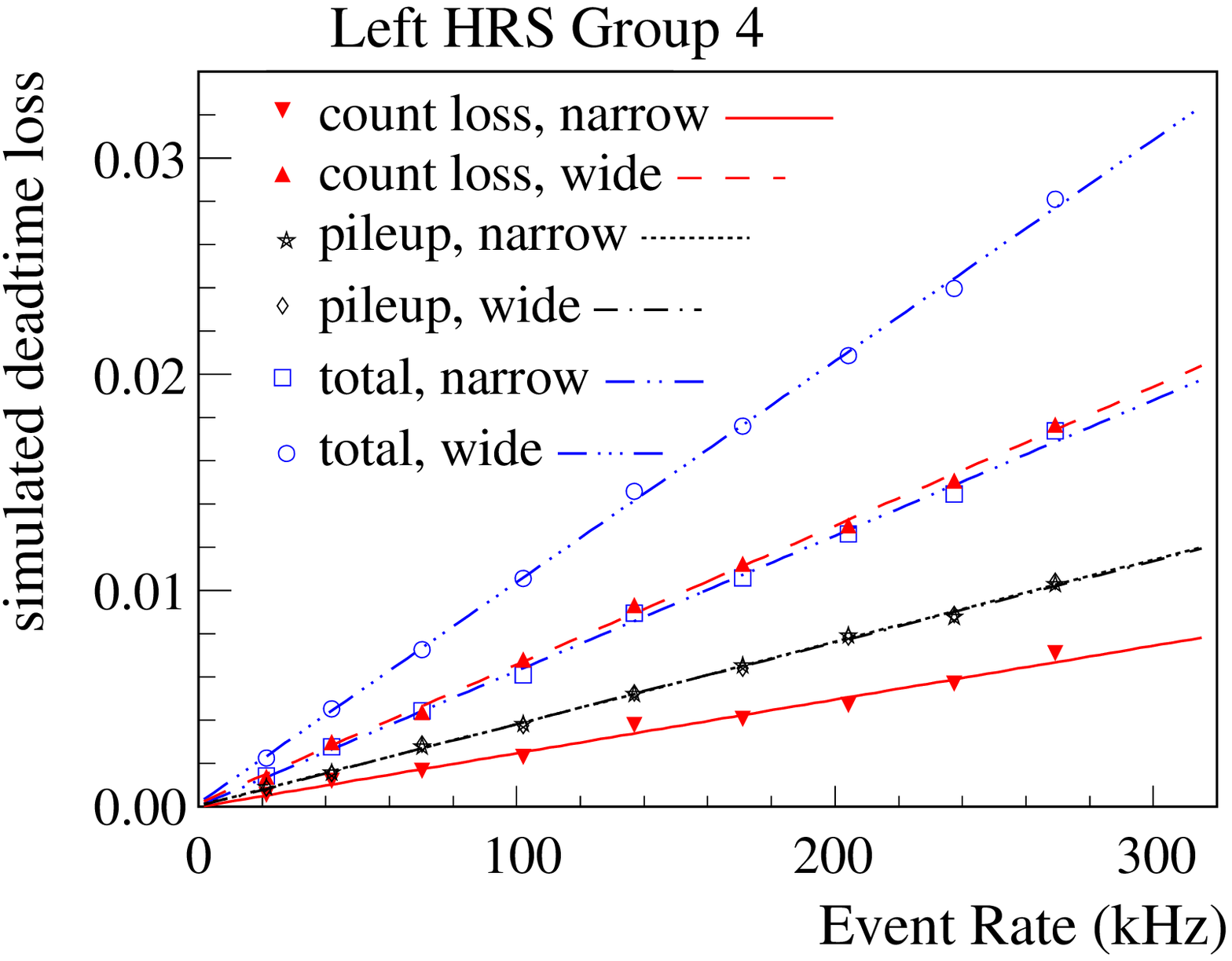}
%black&white version
%  \includegraphics[width=0.6\textwidth,angle=0]{Fig6_1_edit2_bw.eps}
%  \includegraphics[width=0.6\textwidth,angle=0]{Fig6_2_edit2_bw.eps}
 \end{center}
\caption{[{\it Color online}] Deadtime loss vs. event rate from the tagger method for 
group 4 on the Left HRS. Top: actual deadtime loss from tagger measurements; Bottom: simulated
deadtime loss of the tagger. The tagger fractional count loss $1-R_o/R_i$ (fit by solid and dashed lines) and the pileup 
correction $p$ (fit by dotted and dash-dotted lines) are combined to form the total group deadtime $D$ (fit by dash-double-dotted and dash-triple-dotted lines). 
These data were taken (or simulated) at kinematics DIS \#1. 
To minimize the statistical uncertainty while keeping the computing time reasonable, the simulation used
higher event rates than the tagger measurement. 
The total group deadtime can be determined from the linear fit slope coefficients:
tagger data narrow $(61.5\pm 0.2)\times 10^{-9}$~s, wide $(99.9\pm 0.3)\times 10^{-9}$~s,
simulation narrow $(62.5\pm 1.4)\times 10^{-9}$~s, wide $(102\pm 1.3)\times 10^{-9}$~s.
%Results of the linear fit slope coefficient $p_1$ shows the measured or simulated group deadtime
%in seconds
Group 4 is from 
the central blocks of the lead-glass detector and has the highest rate
among all groups.}
\label{fig:dtplot_tagger_l}
\end{figure}

\begin{figure}[!ht]
 \begin{center}
%color version
  \includegraphics[width=0.6\textwidth,angle=0]{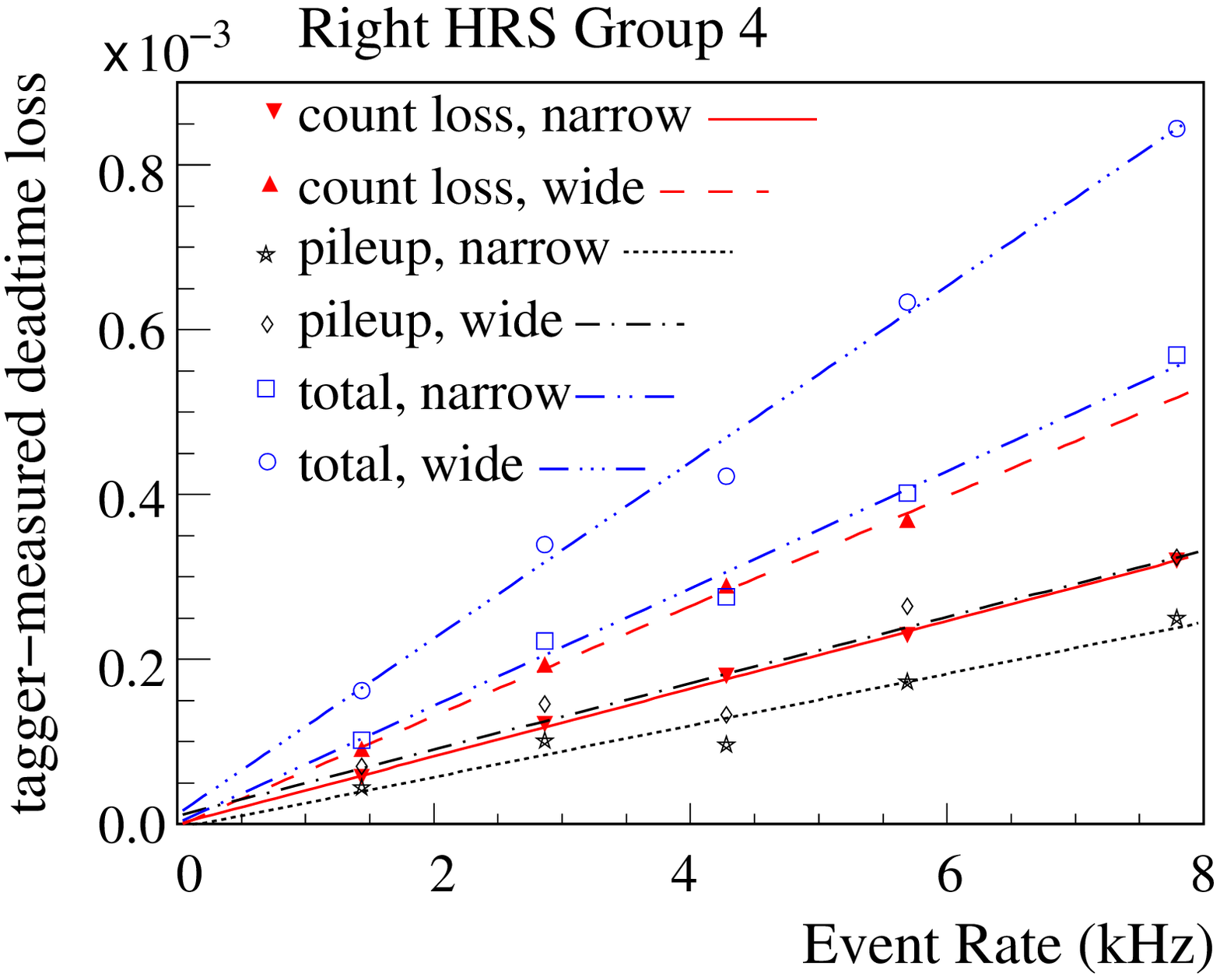}
  \includegraphics[width=0.6\textwidth,angle=0]{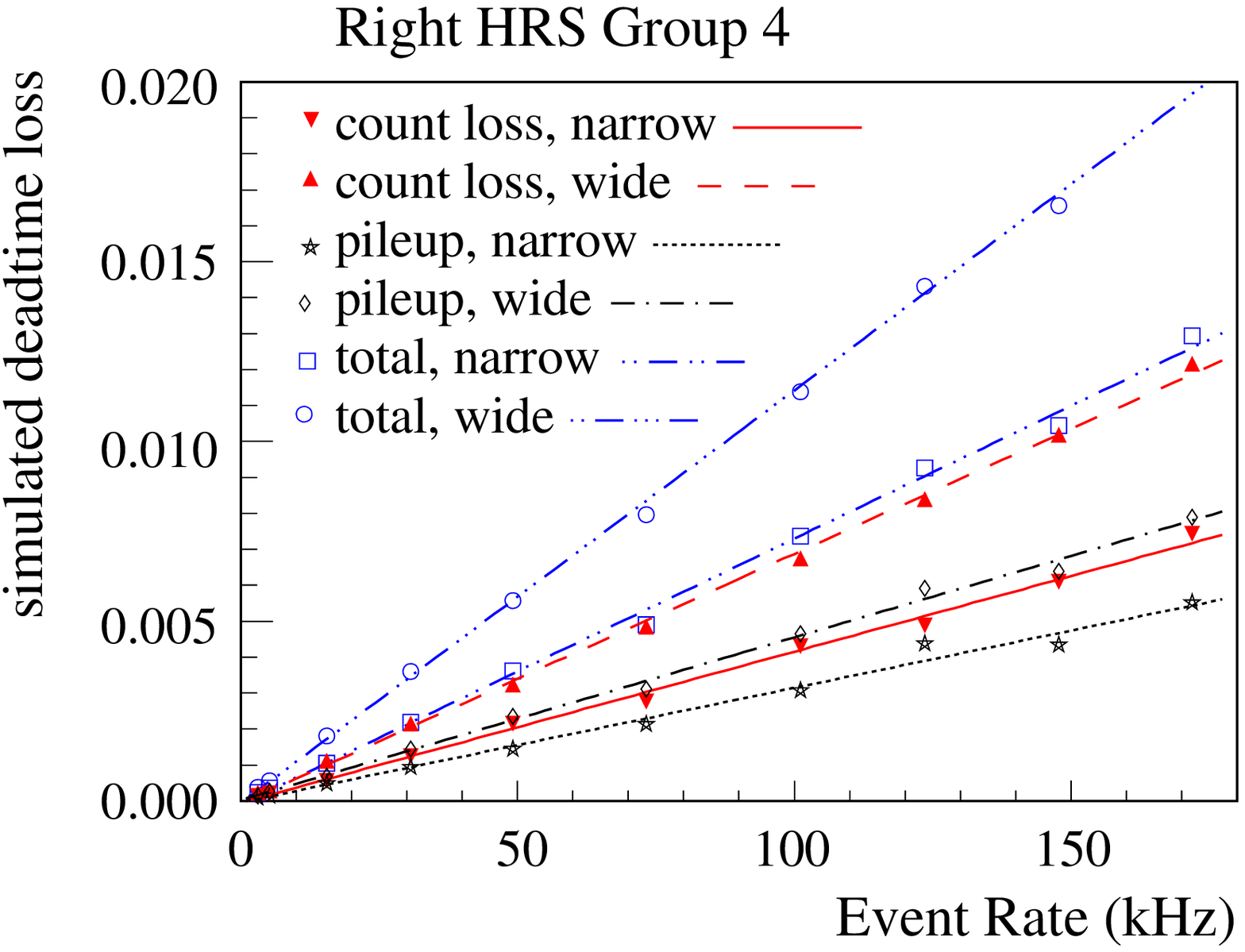}
%black&white version
%  \includegraphics[width=0.6\textwidth,angle=0]{Fig7_1_edit2_bw.eps}
%  \includegraphics[width=0.6\textwidth,angle=0]{Fig7_2_edit2_bw.eps}
 \end{center}
\caption{[{\it Color online}] Deadtime loss vs. group event rate from the tagger method for group 4 on the
Right HRS. Top: tagger data; Bottom: simulation. 
These data were taken (or simulated) at kinematics DIS~\#2.
The total group deadtime can be determined from the linear fit slope coefficients:
tagger data narrow $(71.1\pm 0.9)\times 10^{-9}$~s, wide $(107\pm 1.2)\times 10^{-9}$~s,
simulation narrow $(73.9\pm 1.5)\times 10^{-9}$~s, wide $(115\pm 1.5)\times 10^{-9}$~s.
Group 4 is from 
the central blocks of the lead-glass detector and has the highest rate
among all groups.  See Fig.~\ref{fig:dtplot_tagger_l} caption for more details.}
\label{fig:dtplot_tagger_r}
\end{figure}

The relative loss of tagger events due to DAQ deadtime is evaluated as
\begin{equation}
{D = 1- (1-p)(R_o/R_i),}
\end{equation}
where $R_i$ is the input tagger rate, $R_o$ is the output tagger-trigger coincidence rate,
and $p=(I_1+I_2)/I_0$ is a correction factor for pileup effects as defined in Fig.~\ref{fig:tagger}.
Results for the deadtime loss $D$ are shown in Figs.~\ref{fig:dtplot_tagger_l}
and~\ref{fig:dtplot_tagger_r}, for group 4 on 
the left HRS and group 4 on the right HRS, respectively, and are compared with simulation. 
Different beam currents between 20 and 
100~$\mu$A were used in this dedicated deadtime measurement. In order to reduce the 
statistical fluctuation caused by the limited number of trials in the simulation within a realistic
computing time, simulations were done at higher rates than the actual measurement.

%%%%%%%%%%%%%%%%

\vspace*{0.1cm}

%%%%%%%%%%%%%%%%

The slope of the tagger loss vs. event rate, as shown in Figs.~\ref{fig:dtplot_tagger_l} 
and \ref{fig:dtplot_tagger_r}, gives the value of group deadtime in seconds.
One can see that the deadtime for the wide path is 
approximately 100~ns as expected. The deadtime for the narrow path, on the other hand,
is dominated by the input PMT signal width (typically 60-80~ns) instead of the 30-ns
discriminator width. The simulated group deadtime agrees with the data at a 10\% level or better, 
for both HRSs and for both wide and narrow paths. 

The above tagger measurements were performed at kinematics DIS\#1 on the Left and DIS\#2
on the Right HRS. No tagger data was available for resonance kinematics. However since 
the group deadtime is expected to rely only on the signal width and the module width settings,
as demonstrated by the tagger data, a 10\% systematic uncertainty was used for group deadtime
for all kinematics.

\subsection{Gate Deadtime Evaluation}\label{sec:deadtime_gate}

\begin{figure}
 \begin{center}
 \includegraphics[width=0.8\textwidth]{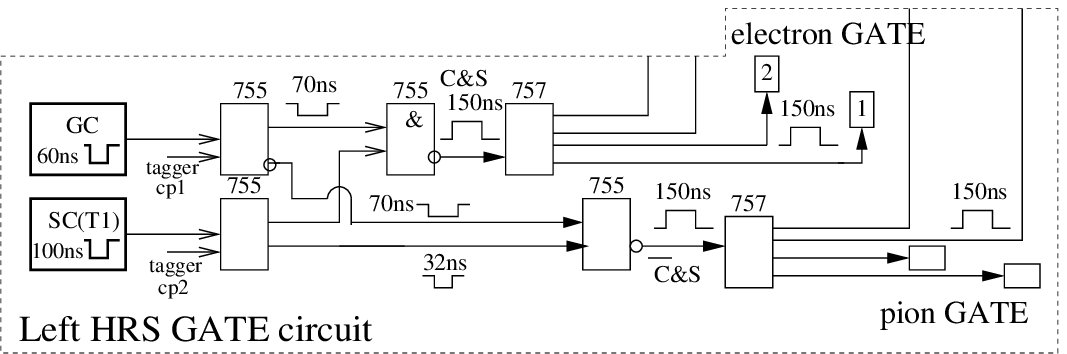}
\vspace*{0.3cm}
 \includegraphics[width=0.8\textwidth]{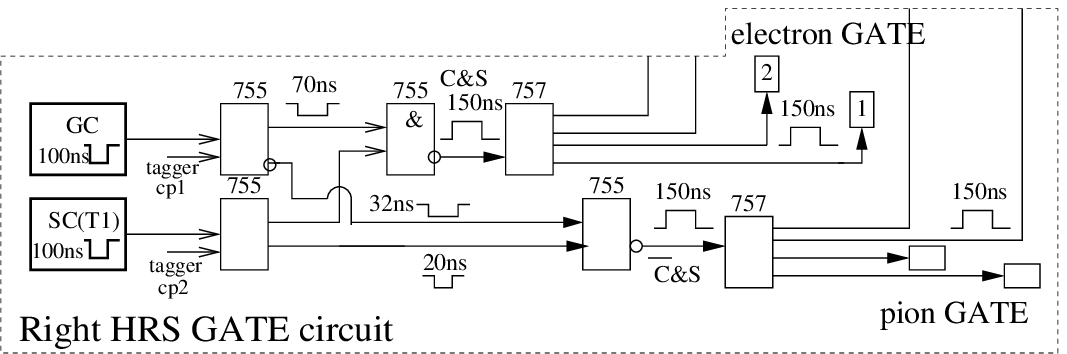}
 \end{center}
 \caption{A zoom-in view of the GATE electronics for Left (top) and Right (bottom) 
spectrometers. }\label{fig:gate}
\end{figure}

Figure~\ref{fig:gate} shows the GATE electronics for both spectrometers, with 
the bottom panel reproducing the GATE portion of Fig.~\ref{fig:daqflowchart}.
It contributes to the total deadtime as follows: 
When both the gas Cherenkov and the Scintillator are triggered by electrons, 
the two signals 
align in time and produce an electron GATE signal. However the Scintillator can be triggered
by pions and other backgrounds, most of which do not trigger the Cherenkov. If an 
electron event arrives shortly after such background events, it triggers the Cherenkov but 
may not trigger the PS755 module that first processes the Scintillator signal because
of the non-updating feature of PS755. In this case, the Cherenkov signal triggered
by the electron may miss the Scintillator signal from the previous pion or background event
and will not produce a valid electron GATE signal. 
Likewise, if a background event triggers the Cherenkov but not the Scintillator, it would also 
cause a loss to the electron events that follow shortly after.  
The fractional loss due to GATE deadtime can be estimated as
\begin{eqnarray}
  DT_\mathrm{gate} &=& R_{SC\&\overline{GC}}(w_{SC,in}-w_{SC,out})+R_{GC\&\overline{SC}}(w_{GC,in}-w_{GC,out}),
 \label{eq:gate_dt}
\end{eqnarray}
where $R_{SC\&\overline{GC}}(R_{GC\&\overline{SC}})$ refers to the rate of events that triggered the Scintillator
(Cherenkov) but not the Cherenkov (Scintillator), $w_{SC,in(out)}$ and 
$w_{GC,in(out)}$ refer to the input (output) signal widths of the PS755 module 
that 
first processes the Scintillator and the Cherenkov signals in the GATE
electronics, respectively. 
Note that if the electronics used to generate the Scintillator and the Cherenkov signals
have intrinsic deadtimes themselves that are longer than $w_{SC,in}$ and $w_{GC,in}$, 
these intrinsic deadtimes should be used in place of the measured $w_{SC,in}$ and $w_{GC,in}$. 
In Eq.~(\ref{eq:gate_dt}), each term on the right
hand side is present only if $w_{in}>w_{out}$. 
From Fig.~\ref{fig:gate}, the signal widths were measured to be: 
$w_{SC,in,\mathrm{left}} = w_{SC,in,\mathrm{Right}}=100$~ns,
$w_{SC,out,\mathrm{left}} = w_{SC,out,\mathrm{Right}} = 32$~ns, 
$w_{GC,in,\mathrm{left}} = 60$~ns,
$w_{GC,in,\mathrm{Right}} = 100$~ns,
$w_{GC,out,\mathrm{left}} = w_{GC,out,\mathrm{Right}} = 70$~ns.
However it was observed from the data that the Left HRS Cherenkov signal had an intrinsic
deadtime of longer than 70~ns. In fact, data showed both HRSs had contributions from the two terms
on the right hand side of Eq.~(\ref{eq:gate_dt}). 

Because trigger rates from Scintillator and the gas Cherenkov were much higher than 
individual group rates, the GATE deadtime could dominate the total deadtime of the DAQ, 
and the difference in total deadtime loss between narrow and wide paths could be smaller than
that in their group deadtimes. 

\begin{figure}[!htp]
%\begin{center}
%color version
\includegraphics[width=0.5\textwidth]{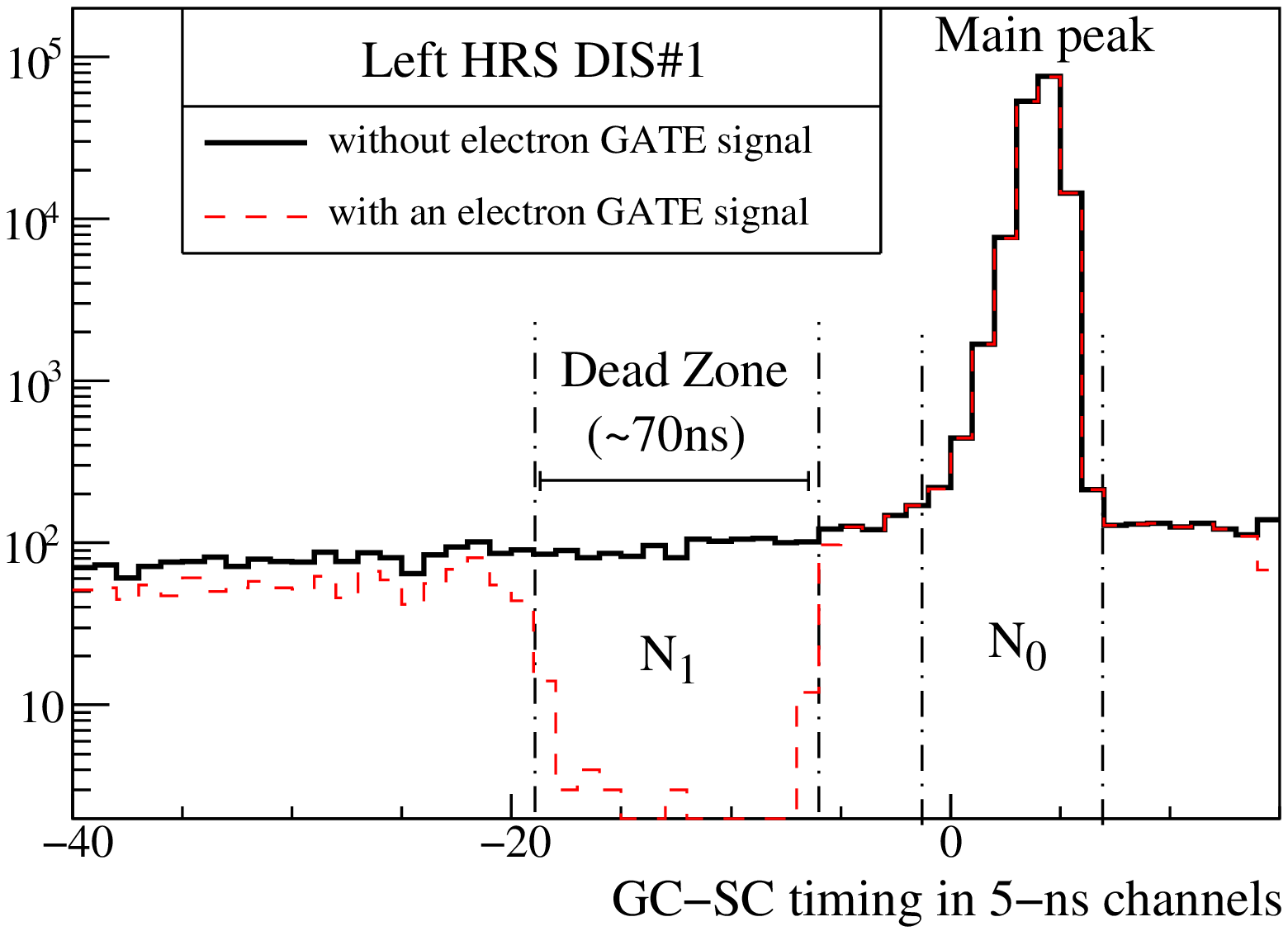}
\includegraphics[width=0.5\textwidth]{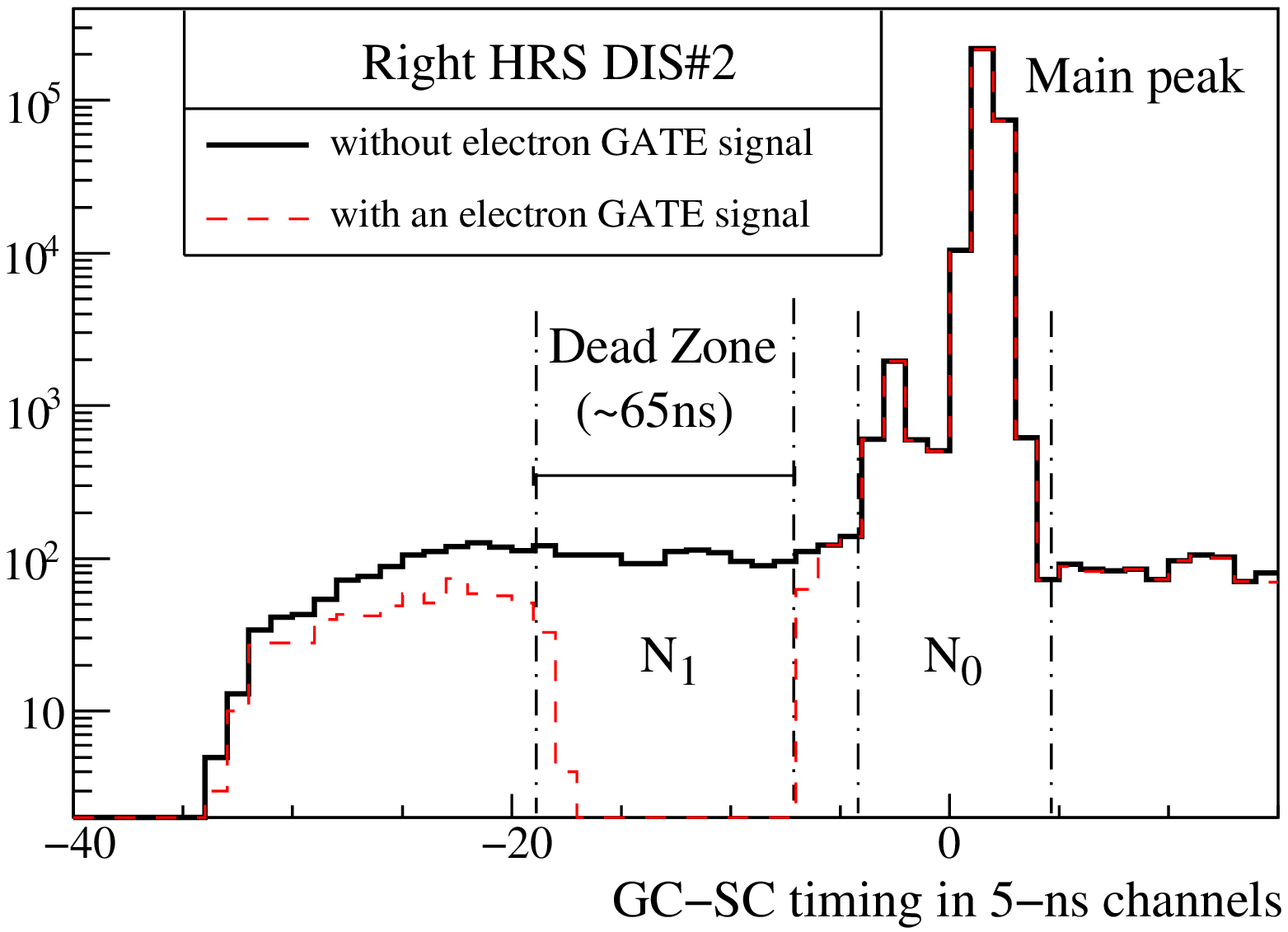}
%black&white version
%\includegraphics[width=0.5\textwidth]{veto_tdc_left_edit_bw.eps}
%\includegraphics[width=0.5\textwidth]{veto_tdc_right_edit_bw.eps}
%\end{center}
\caption{[{\it Color online}] Timing difference between Gas Cherenkov and Scintillator signals in 5-ns channels. These data were taken with a beam current of 
110~$\mu$A and at kinematics DIS\#1 on the Left and DIS\#2 on the Right HRS, respectively. The fractional loss of electron events due to GATE deadtime can be estimated using the ratio of $N_1/N_0$, where $N_1$ is the count difference between the two spectra in the dead zone, and $N_0$ is the counts under the main peak near 0~ns.  See text for details.
}
\label{fig:gate_fadc}
\end{figure}

The GATE deadtime can be extracted from the trigger simulation HATS using the known signal widths
and module settings, and be compared with the estimation of Eq.~(\ref{eq:gate_dt}). In addition,
evidence of the GATE deadtime can be extracted from FADC data. 
Figure~\ref{fig:gate_fadc} shows spectra of the timing difference between the gas Cherenkov (GC) 
and the Scintillator (SC) signals extracted from FADC data. Timing of the GC signal should represent 
the timing of an electron event, while the SC signal can be triggered by the same electron 
(as represented by the main peak near 0~ns), or a pion event that preceded the electron (as represented by the region
$<0$~ns). The region beyond $\pm 100$~ns were pure random events since the SC signal input to the GATE
electronics was only 100~ns wide.
As one can see, the region between 
-100 and $\approx -30$~ns represents a ``dead zone'' where the preceding pion triggered the PS755 unit
that first processed the SC signal, and caused the electron events that followed to not trigger the GATE
circuit.  The probability for the electron events to not be recorded by the DAQ due to this GATE deadtime
is thus the ratio of the dead zone area ($N_1$) and the area of the main peak near 0~ns ($N_0$), 
see Fig.~\ref{fig:gate_fadc}.

\begin{figure}[!htp]
%\begin{center}
%color version
%\includegraphics[width=0.5\textwidth]{dt_veto_left_new_edit.eps}
%directly from DW
\includegraphics[width=0.5\textwidth]{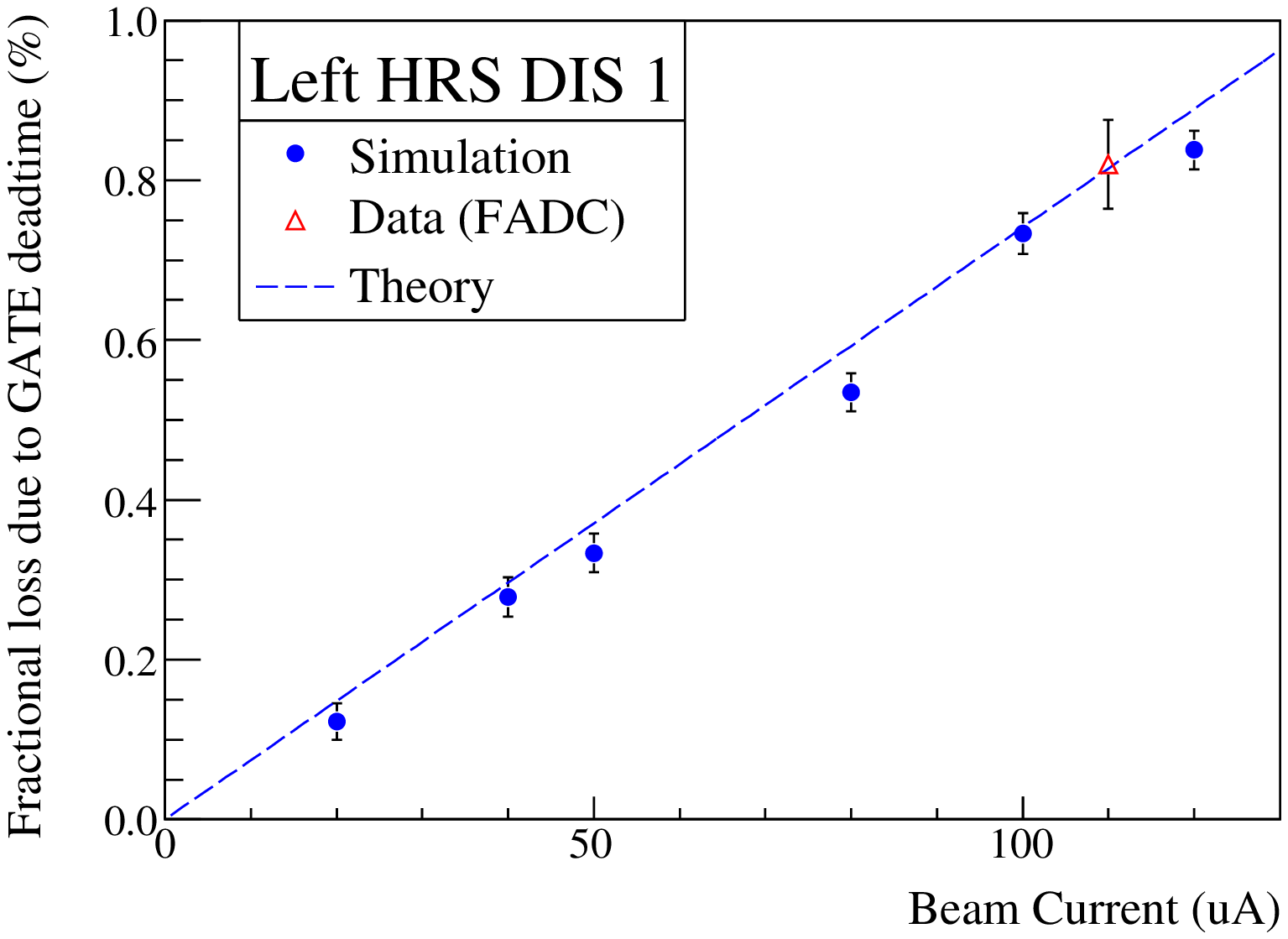}
\includegraphics[width=0.5\textwidth]{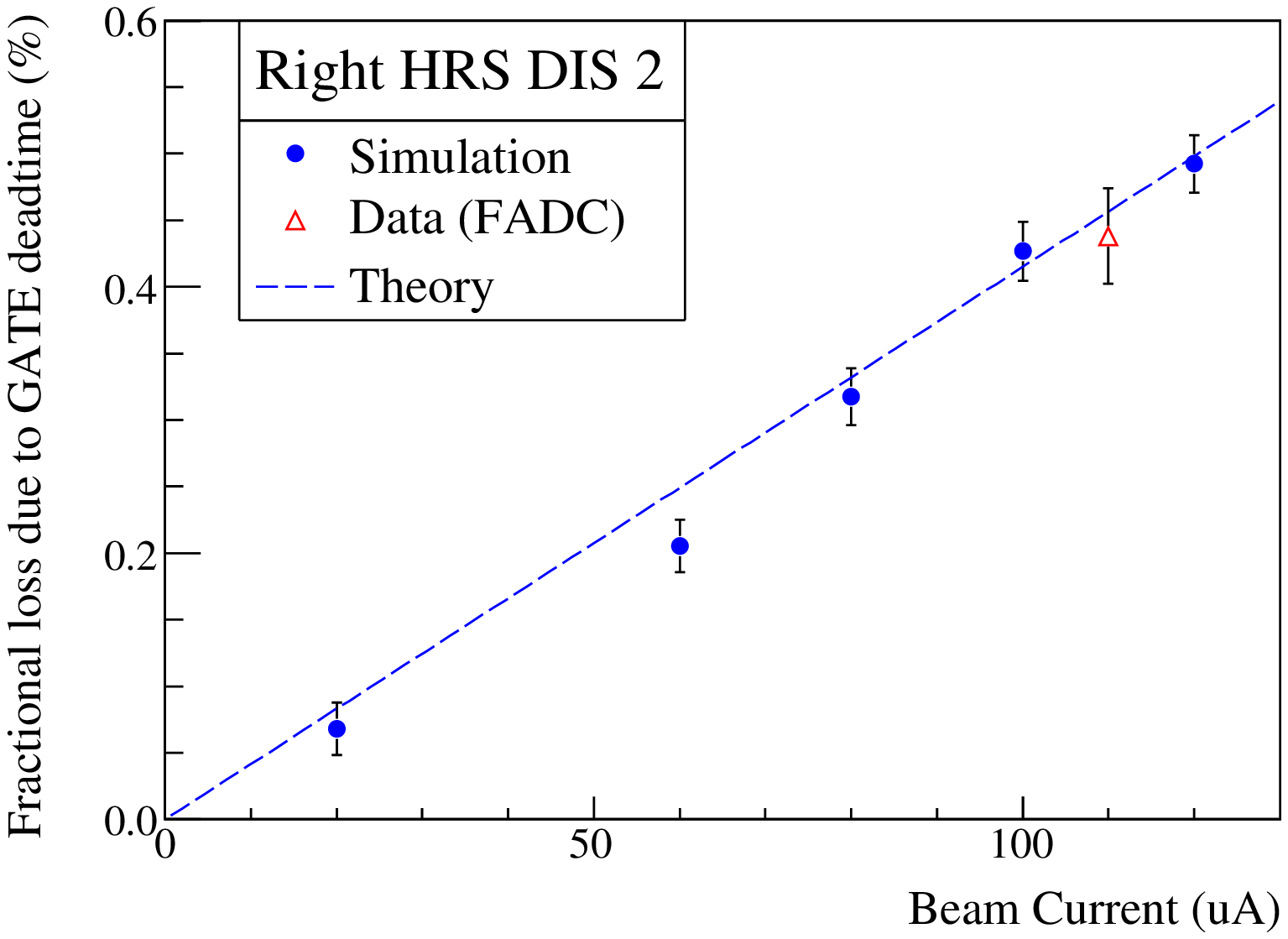}
%black&white version
%directly from DW
%\includegraphics[width=0.5\textwidth]{dt_veto_left_bw.eps}
%\includegraphics[width=0.5\textwidth]{dt_veto_righta_bw.eps}
 \caption{[{\it Color online}] Fractional loss due to GATE deadtime as a function of beam current 
obtained from trigger simulation (solid circles), the analytic method Eq.~(\ref{eq:gate_dt})
(dashed line), and FADC data (open triangle). Error bars for simulations are statistical.}
\label{fig:gatedt_comp}
\end{figure}

Figure~\ref{fig:gatedt_comp} shows comparisons of the fractional losses due to GATE deadtime estimated using 
 trigger simulation, the analytic method Eq.~(\ref{eq:gate_dt}), and FADC data extracted from
Fig.~\ref{fig:gate_fadc}.
The agreement between simulation and FADC was found to be better than 10\% and this was used 
as the systematic uncertainty of the GATE deadtime. 
For resonance kinematics no FADC data was available. GATE deadtime for resonance data was obtained from trigger 
simulation and the same systematic uncertainty
was used because the mechanism of the GATE deadtime was expected to remain the same throughout the
experiment.

\subsection{OR Deadtime Evaluation}\label{sec:deadtime_or}

There is no direct measurement of the logical OR deadtime, but the effect of the logical OR
module is straightforward and can be calculated analytically: When
two electron triggers from different groups overlap in time as they arrive at the logical
OR module, they generate only one output in the global trigger. 
This OR deadtime loss can be calculated using the recorded trigger rates and the
known trigger signal widths. 
To confirm the analytic method results, the OR deadtime was evaluated from trigger simulation
by subtracting the group and the GATE deadtimes from the total deadtime, all three of
which were direct results from the simulation. 
%Comparisons between the simulation and the analytic method are shown in
%Table~\ref{tab:ORdt}, together with the systematic error calculated using the difference of the two. 
The difference between the analytic method and trigger simulation was used as the systematic uncertainty of the OR
deadtime. 
%The fact that the OR deadtime only contributes a small fraction to the total deadtime makes it less demanding to know it precisely.

\subsection{Total Deadtime Evaluation}\label{sec:deadtime_final}

\begin{figure}[!ht]
%\begin{center}
%color version
 \includegraphics[width=0.51\textwidth,angle=0]{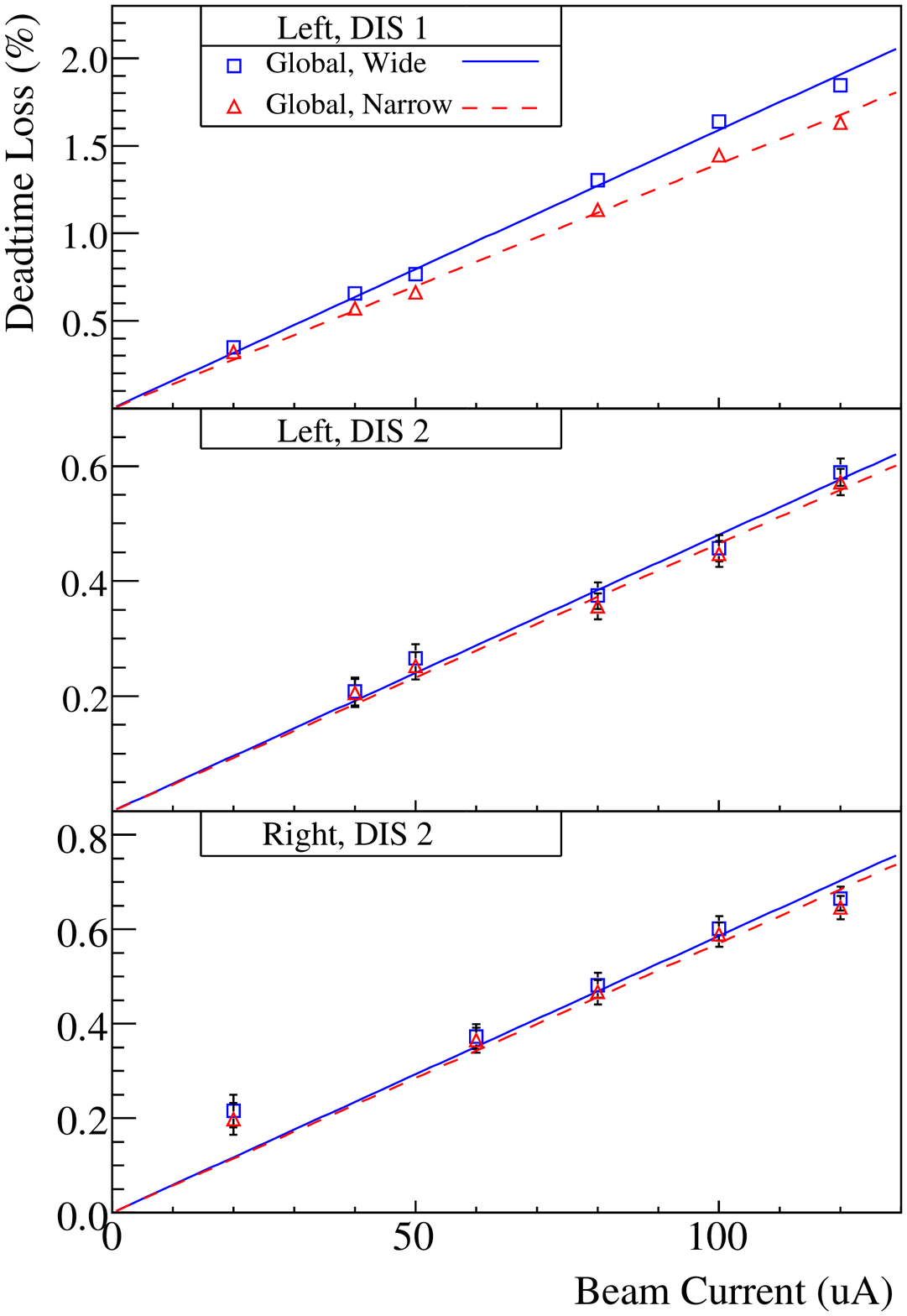}
 \includegraphics[width=0.49\textwidth,angle=0]{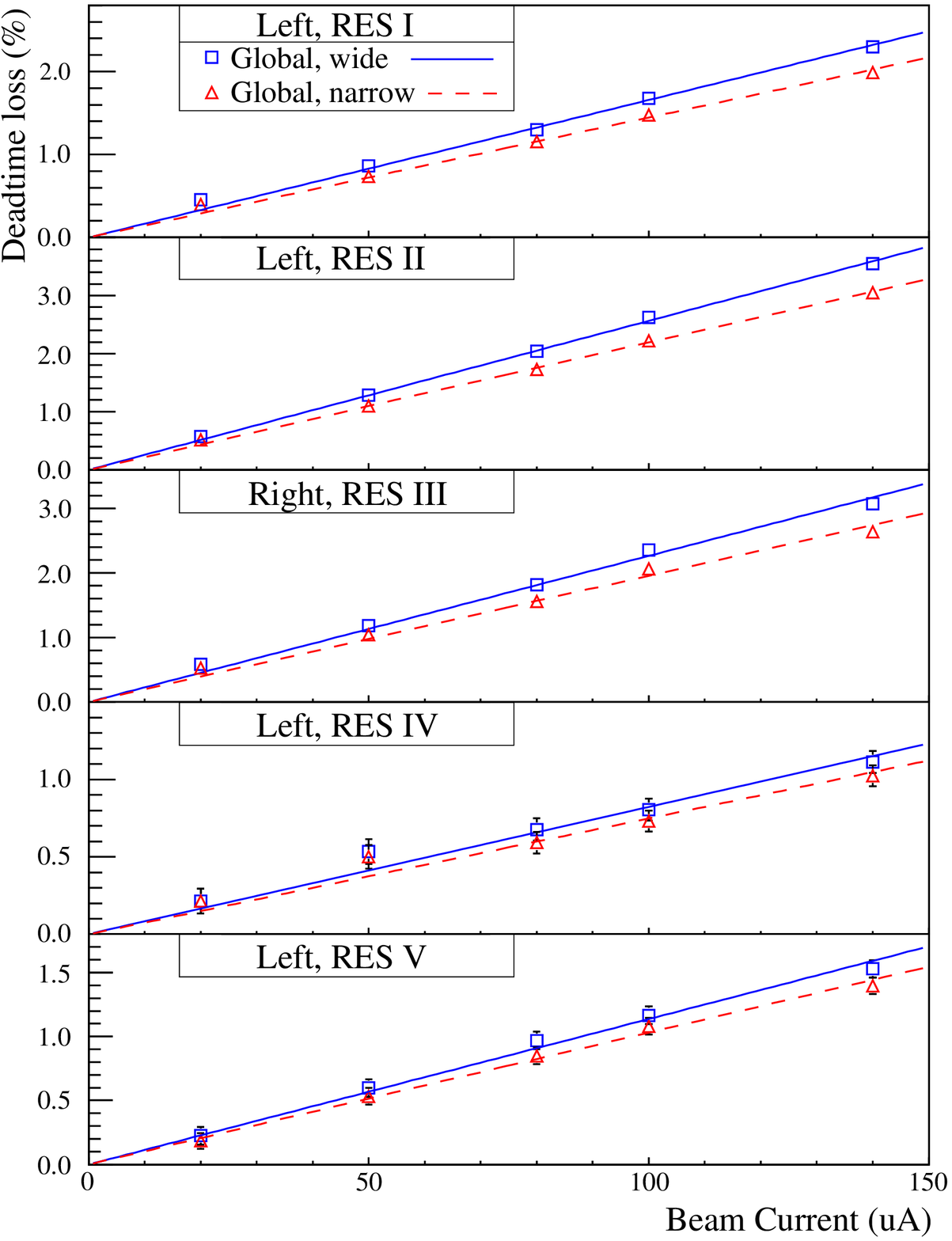}
%black&white version
%directly from DW
% \includegraphics[width=0.51\textwidth,angle=0]{dtall_bw.eps}
% \includegraphics[width=0.49\textwidth,angle=0]{dtres_edit_bw.eps}
%\end{center}
\caption{[{\it Color online}] Simulated deadtime loss of the global electron trigger for
the three DIS spectrometer and kinematics combinations and the five resonance kinematics, 
for the narrow path (open triangles, fit by dashed lines) and the wide path (open squares, fit
by solid lines). 
The error bars shown are due to statistical
uncertainty of the simulation. 
See Table~\ref{tab:deadtime_all} for final uncertainty evaluation of the total deadtime loss.
}\label{fig:hats_dtoverall}
\end{figure}

The simulated deadtime loss of the global electron triggers and its decomposition
into group, GATE, and OR are shown in Table~\ref{tab:deadtime_all}, along with the 
total deadtime correction at a beam current of 100~$\mu$A. 
The total deadtime loss not only increases with higher electron rate $R_e$, but also with
higher pion to electron ratio $R_\pi/R_e$ (see Table~\ref{tab:kine}) which would cause larger GATE deadtime.
The deadtime loss 
is also shown in Fig.~\ref{fig:hats_dtoverall} as a function of the total event rate.
\begin{table}[!htp]
 \caption{Simulated DAQ deadtime loss in percent for all kinematics and for both narrow (n) 
and wide (w) paths, along with the fractional contributions from group, GATE, and OR deadtimes. 
The
fractional deadtime from OR was calculated as one minus those from group and GATE, and 
its uncertainty was estimated from the difference between simulation and the analytical results. 
The variation of group deadtime contribution among kinematics is due to changes in the rate distribution among different groups.
The uncertainty of the total deadtime is the uncertainties from group, GATE and OR added
in quadrature.
}\label{tab:deadtime_all}
 \begin{center}
 \begin{tabular}{c|c|c|c|c|c}
  \hline\hline
  Kine,    & Path  & \multicolumn{3}{c|}{fractional contribution} & Total deadtime \\
 \cline{3-5}
  HRS      &       & Group           & GATE   & OR  & loss at 100$\mu$A  \\ \hline
% 
%             & n &$(20.6\pm 2.1)\%$&$(51.3\pm 1.9)\%$ &$(28.1\pm 8.6)\%$& $(1.45\pm 0.13)\%$  \\
%             & w &$(29.5\pm 2.4)\%$&$(45.3\pm 1.7)\%$ &$(25.3\pm 9.0)\%$ & $(1.64\pm 0.16)\%$ \\
%VETO error scaled by 2.372 to account for stat, OR dt error corrected
   DIS\#1, & n &$(20.6\pm 2.1)\%$&$(51.3\pm 3.5)\%$ &$(28.1\pm 4.7)\%$& $(1.45\pm 0.09)\%$  \\
   Left    & w &$(29.5\pm 2.4)\%$&$(45.3\pm 3.1)\%$ &$(25.3\pm 4.6)\%$ & $(1.64\pm 0.10)\%$ \\
  \hline
 % For Left kine 2 veto, no FADC available and used fractional error of left kine 1.
   DIS\#2, & n &$(5.4\pm 0.8)\%$ &$(81.1\pm 5.5)\%$&$(13.5\pm 7.0)\%$ & $(0.50\pm 0.04)\%$\\
   Left    & w &$(8.4\pm 0.4)\%$ &$(77.3\pm 5.3)\%$&$(14.3\pm 8.0)\%$ & $(0.52\pm 0.05)\%$\\
  \hline
   DIS\#2, & n &$(4.6\pm 0.4)\%$ &$(72.9\pm 6.0)\%$&$(22.6\pm 17.4)\%$ & $(0.57\pm 0.10)\%$\\
   Right   & w &$(6.9\pm 0.7)\%$ &$(71.0\pm 5.8)\%$&$(22.1\pm 17.9)\%$ & $(0.58\pm 0.11)\%$\\
  \hline
   RES I,  & n & $(26.3\pm 3.8)\%$ &$(39.3\pm 2.7)\%$&$(34.4\pm 1.8)\%$ & $(1.45\pm 0.07)\%$\\
   Left     & w &$(37.2\pm 2.1)\%$ &$(34.3\pm 2.3)\%$&$(28.5\pm 3.1)\%$ & $(1.66\pm 0.07)\%$\\
  \hline
   RES II, & n & $(27.6\pm 4.3)\%$ &$(38.8\pm 2.7)\%$&$(33.6\pm 7.5)\%$ & $(2.19\pm 0.20)\%$\\
   Left   & w &$(38.3\pm 1.9)\%$ &$(33.2\pm 2.3)\%$&$(28.5\pm 7.0)\%$ & $(2.56\pm 0.19)\%$\\
  \hline
   RES III, & n & $(22.9\pm 1.8)\%$ &$(60.0\pm 4.9)\%$&$(17.1\pm 18.48)\%$ & $(1.96\pm 0.38)\%$\\
   Right     & w &$(30.8\pm 3.1)\%$ &$(51.8\pm 4.3)\%$&$(17.4\pm 12.73)\%$ & $(2.27\pm 0.31)\%$\\
  \hline
   RES IV,  & n & $(14.5\pm 1.9)\%$ &$(63.7\pm 4.4)\%$&$(21.9\pm 3.0)\%$ & $(0.75\pm 0.04)\%$\\
   Left     & w &$(21.5\pm 1.0)\%$ &$(58.2\pm 4.0)\%$&$(20.3\pm 2.9)\%$ & $(0.82\pm 0.04)\%$\\
  \hline
   RES V,   & n & $(15.5\pm 2.1)\%$ &$(68.3\pm 4.7)\%$&$(16.2\pm 5.7)\%$ & $(1.03\pm 0.08)\%$\\
   Left     & w &$(22.7\pm  1.1)\%$ &$(61.7\pm 4.2)\%$&$(15.6\pm 3.0)\%$ & $(1.14\pm 0.06)\%$\\
\hline\hline
 \end{tabular}
 \end{center}
\end{table}
% from top to bottom: 9.0, 9.5, 22.2, 23.1%

Results shown in Table~\ref{tab:deadtime_all}
provide a direct correction to the measured asymmetry, and the uncertainties
are small compared with other dominant systematic uncertainties such as the 
approximately 2\% uncertainty from beam polarizations.
In practice, the deadtime correction was applied to data on a run-by-run basis with the deadtime of each run calculated using the actual beam current during the run and the linear fitting results from Fig.~\ref{fig:hats_dtoverall}.

%%%%%%%%%%%%%%%%
\subsection{Asymmetry Measurement}
%%%%%%%%%%%%%%%%
The physics asymmetries sought for in this experiment were expected to be in the order of $10^2$~ppm. 
The measured asymmetries were about $90\%$ of the expected values due to beam polarization.
To understand the systematics of the asymmetry measurement, a half-wave plate (HWP) was
inserted in the beamline to flip the laser helicity in the polarized source during half
of the data taking period. The measured asymmetries flipped sign for each beam HWP change
and the magnitude of the asymmetry remained consistent within statistical error bars. 

\begin{figure}[!ht]
 \includegraphics[width=0.5\textwidth,angle=0]{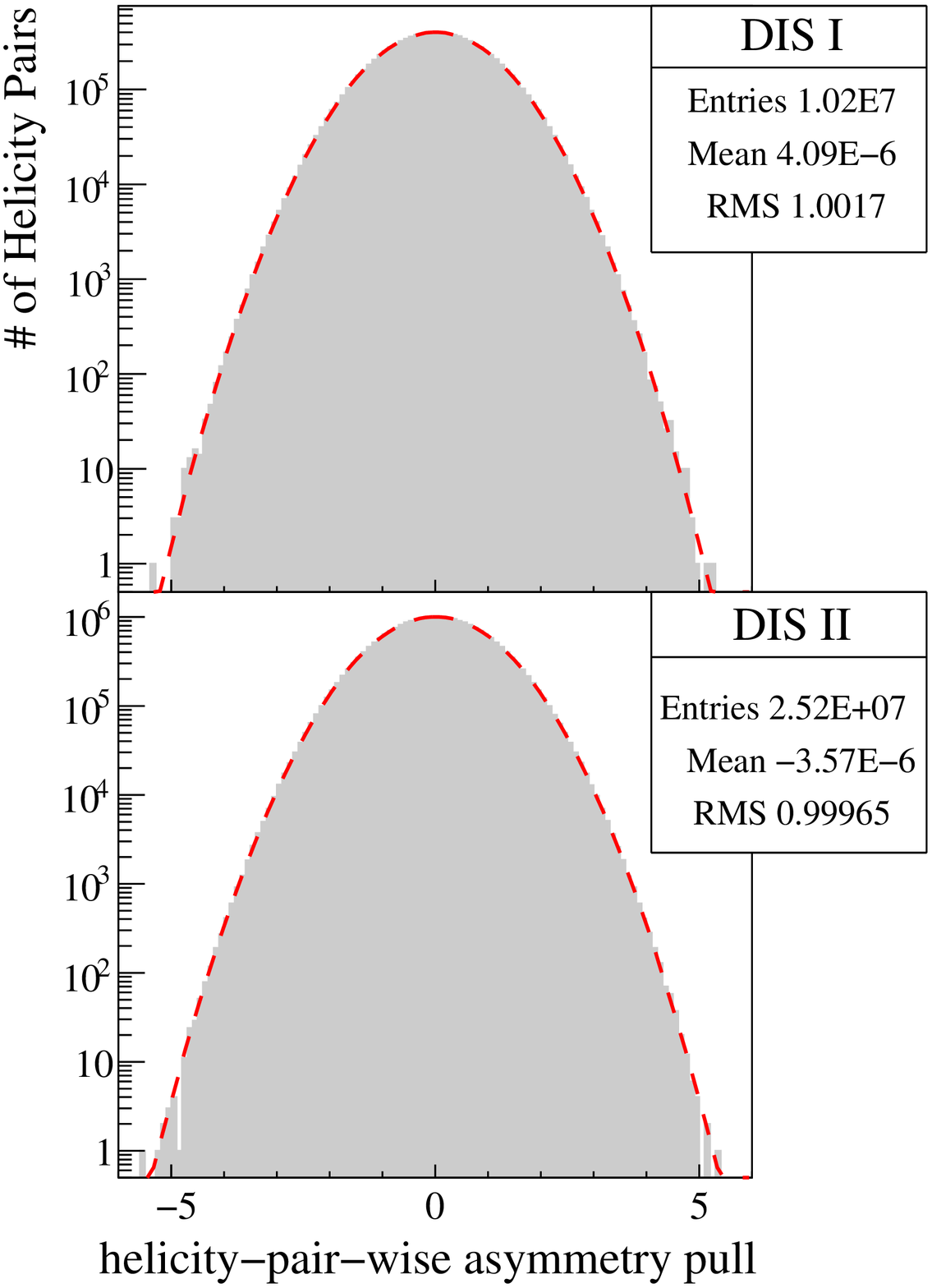}
 \includegraphics[width=0.5\textwidth,angle=0]{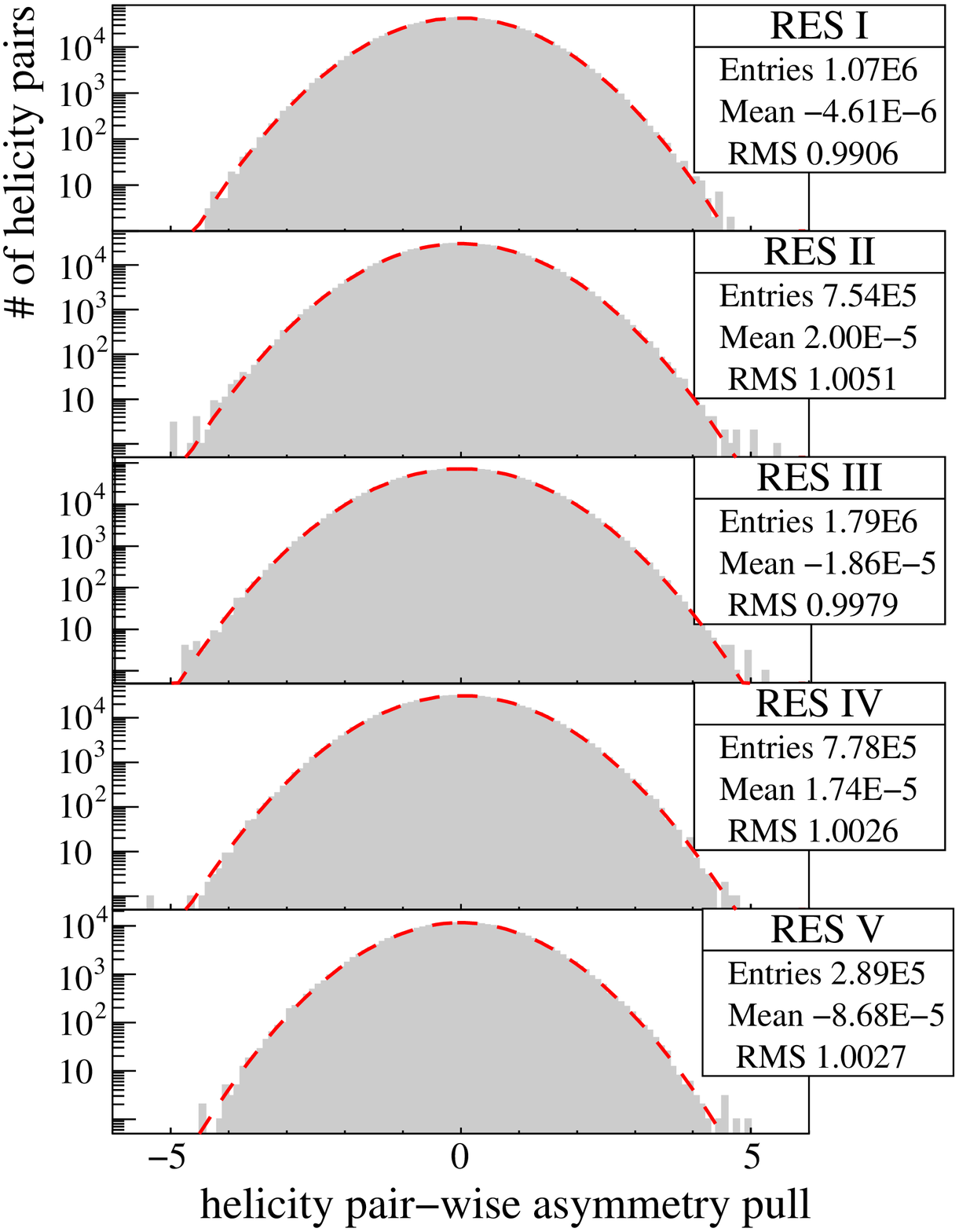}
%black&white version
% \includegraphics[width=0.5\textwidth,angle=0]{pair_pull_DIS_edit_bw.eps}
% \includegraphics[width=0.5\textwidth,angle=0]{pair_pull_RES_edit_bw.eps}
\caption{[{\it Color online}] Pull distribution~[Eq.(\ref{eq:pull})] for the global electron 
narrow trigger for the three DIS spectrometer and kinematics combinations, and the five 
resonance kinematics. The DIS\#2 distribution has both the Left and the Right HRS
data combined.
}\label{fig:asym}
\end{figure}

The asymmetries can be formed from event counts of each beam helicity pair, with 
33-ms of helicity right and 33-ms of helicity left beam, normalized by the beam charge.
Figure~\ref{fig:asym} shows the pull distribution of these pair-wise asymmetries with the ``pull'' 
defined as
\begin{eqnarray}
 p_i &\equiv& (A_i-\langle A\rangle)/{\delta A_i}~,~\label{eq:pull}
\end{eqnarray}
where $A_i$ is the asymmetry extracted from the $i$-th beam helicity pair with the HWP
states already corrected and
$\delta A_i=1/\sqrt{N_i^R+N_i^L}$ its statistical uncertainty with $N_i^{R(L)}$ the
event count from the right (left) helicity pulse of the pair, and $\langle A\rangle$
is the asymmetry averaged over all beam pairs.  One can see that the asymmetry spectrum agrees to 
five orders of magnitude with the Gaussian distribution, as expected from purely statistical
fluctuations.

%%%%%%%%%%%%%%%%

%%%%%%%%%%%%%%%%
\section {Summary}
%%%%%%%%%%%%%%%%
A scaler-based counting DAQ with hardware-based particle identification 
was successfully implemented in the 6 GeV 
PVDIS experiment at Jefferson Lab to measure parity-violating asymmetries
at the $10^{-4}$ level at event rates of up to 600~kHz. 
Asymmetries measured by the DAQ followed
Gaussian distributions as expected from purely statistical measurements.
Particle identification performance of the DAQ was 
measured and corrections were applied to the data on a day-to-day 
basis. The overall pion contamination in the electron sample was controlled
to approximately $2\times 10^{-4}$ or lower, with an electron
efficiency above 91\% during most of the data production period of the experiment. 
The DAQ deadtime was evaluated from a full-scale timing simulation and
contributed an uncertainty of no more than $0.5\%$ to the final asymmetry results.
Systematic uncertainties from the pion contamination and the counting deadtime therefore were
both negligible compared to the $(3-4)\%$ statistical uncertainty and other 
leading systematic uncertainties. 
Results presented here demonstrate that accurate asymmetry measurements can be performed
with even higher event rates or backgrounds with this type of scaler-based DAQ. 

%%%%%%%%%%%%%%%%
\section*{Acknowledgments}

This work was supported in part by the Jeffress Memorial Trust under Award No. J-836, 
the U.S. National Science Foundation under Award No. 0653347, 
and the U.S. Department of Energy under Award No. DE-SC0003885 
and DE-AC02-06CH11357.
{\bf Notice:} Authored by Jefferson Science Associates, LLC under U.S. DOE
Contract No. DE-AC05-06OR23177. The U.S. Government retains a non-exclusive, paid-up,
irrevocable, world-wide license to publish or reproduce this manuscript for U.S.
Government purposes.
%%%%%%%%%%%%%%%%

%%%%%%%%%%%%%%%%
%\input pvdis_nim_bib

%%%%%%%%%%%%%%%%
\end{document}